\documentclass[12pt]{article}
\usepackage{euscript,amsfonts,amsmath,amssymb,color}
\usepackage{graphicx}
\usepackage{caption}
\usepackage{soul}
\usepackage{subcaption}
\usepackage{fullpage}
\usepackage[round,authoryear]{natbib}
\usepackage{array}
\usepackage{setspace}
\usepackage{soul}

\definecolor{webgreen}{rgb}{0,0.4,0}
\definecolor{webbrown}{rgb}{0.6,0,0}
\definecolor{purple}{rgb}{0.5,0,0.25}
\definecolor{darkblue}{rgb}{0,0,0.7}
\definecolor{darkred}{rgb}{0.7,0,0}
\definecolor{darkgreen}{rgb}{0,0.7,0}
\usepackage[pdfborder=false]{hyperref}
\hypersetup{colorlinks,citecolor=darkred,filecolor=black,linkcolor=darkblue,urlcolor=webgreen,pdfpagemode=none,pdfstartview=FitH}
\newcommand{\ignore}[1]{}
\newtheorem{lemma}{{\bf Lemma}}
\newtheorem{proposition}{{\bf Proposition}}
\newtheorem{corollary}{{\bf Corollary}}
\newtheorem{theorem}{{\bf Theorem}}
\newtheorem{defn}{{\bf Definition}}

\newtheorem{example}{{\bf Example}}

\newcommand{\marginnote}[1]{\marginpar{\tiny\raggedright#1}}

\newcommand{\bikh}[1]{\marginnote{\textcolor{blue}{{\bf Bikh: }#1}}}
\newcommand{\debasis}[1]{\marginnote{\textcolor{webbrown}{{\bf Debasis: }#1}}}
\newcommand{\aed}[1]{{\color{black} #1}}
\newcommand{\rfr}[1]{{\color{black} #1}}

\def\uv2{\underline{v}_2}
\def\ov2{\overline{v}_2}

\def\grad{\bigtriangledown}
\def\e{\mbox{\rm E}}

\newcommand{\rarely}[1]{}

\begin{document}

\allowdisplaybreaks

\begin{titlepage}
\title{\textbf{Selling Two Identical Objects}\thanks{We are grateful to Moshe Babaioff, Tilman B{\"o}rgers, Rahul Deb, Bhaskar Dutta, Jingtao Li, Aroon Narayanan, Kolagani Paramahamsa, two anonymous referees,  an Associate Editor, and seminar participants at Ashoka University, Delhi Economic Theory Workshop, Essex University,  University of Michigan, Penn State University, Stony Brook University,  University of Toronto, and UCLA for helpful comments. }}
\author{Sushil Bikhchandani\thanks{Anderson School at UCLA, Los Angeles ({\tt sbikhcha@anderson.ucla.edu}).}$\;\;$ and Debasis Mishra\thanks{Indian Statistical Institute, Delhi ({\tt dmishra@isid.ac.in}).}}
\maketitle

\begin{abstract}
It is well-known that optimal (i.e., revenue-maximizing) selling mechanisms in multidimensional type spaces may involve randomization. We obtain conditions under which deterministic mechanisms are optimal for selling two identical, indivisible objects to a single buyer. We analyze two settings: (i)~decreasing marginal values (DMV) and (ii)~increasing marginal values (IMV). Thus, the values of the buyer for the two units are not independent.

We show that under a well-known condition on distributions~(due to \cite{MM88}), (a) it is optimal to sell the first unit deterministically in the DMV model and (b) it is optimal to bundle (which is a deterministic mechanism) in the IMV model. Under a stronger sufficient condition on distributions, a deterministic mechanism is optimal in the DMV model.

Our results apply to heterogeneous objects when there is a specified sequence in which the two objects must be sold.

\bigskip
\noindent
JEL Classification number: D82

\noindent
Keywords: multiobject auctions, revenue maximization, multidimensional mechanism design

\end{abstract}
\thispagestyle{empty}
\end{titlepage}

\section{Introduction}

We consider optimal, i.e., expected revenue maximizing, mechanisms for selling two identical units of an object to a buyer. The buyer's type (values for the units) is two dimensional and privately known to the buyer. We focus on two cases: decreasing marginal values and increasing marginal values. Thus, the buyer's values for the units are not independent.

 A general solution to the optimal mechanism design problem for the sale of multiple indivisible products is unknown. Unlike the single product case, the optimal mechanism for selling two or more products may involve randomization (see \cite{Th04}, \cite{MV06}, \cite{Py06}, and \cite{HR15}).  Our objective is to find sufficient conditions under which a deterministic mechanism is optimal among all mechanisms for selling two identical units, including random mechanisms.

 We assume that the seller can commit to a mechanism. Implicit in this is the assumption  that the mechanism can be objectively verified by both parties. As \citet{LM02} emphasize, it is easier to verify a deterministic mechanism than a random mechanism. For instance, commitment by a seller to a random mechanism may not be credible in a one-shot interaction with a buyer. Perhaps this is a reason for the limited use of randomized selling methods.\footnote{Random selling methods, called opaque selling, are used by travel websites such as Hotwire and Priceline.}
 Under our sufficient conditions, the seller does not sacrifice optimality for credible commitment to a (deterministic) mechanism.

 The assumption of homogeneous objects reduces the dimensionality of the price space and therefore  the dimensionality of random allocation rules (compared to heterogeneous objects). While this represents a simplification of the problem of finding an optimal mechanism, the correlation of values in our paper increases complexity.

With homogeneous objects, there is a natural order of transactions: the second unit can be sold only after the first unit is sold. In some settings with two {\sl heterogeneous} objects, one of the two objects can be sold only after the other object is sold.\footnote{See \cite{Ar16} for a discussion of this issue.} For instance, the warranty on a product is only sold to a buyer who purchases the product. Another example is when a seller offers two versions of a product, basic or premium. The premium version of a product can be viewed as the basic version plus an upgrade. That is, the upgrade is sold only if the basic product is also sold. All our results apply to such settings.

We define a function of buyer (marginal) values,\footnote{As the values of the two units are additive, the value of a unit is the same as its marginal value. } $\Phi(v_1,v_2)$, which plays a key role in the analysis.\footnote{The buyer's values (or type) for the two units are  $v_1\in[0,1]$ and $v_2\in[0,a]$. }  The function $\Phi$ is a guidepost for making revenue improvements to any incentive compatible and individual rational mechanism. If $\Phi$ satisfies certain single-crossing conditions, then incentive compatibility and individual rationality are maintained in the improved mechanism. The function $\Phi$ depends only on the distribution of types.

With decreasing marginal values, we show that if $\Phi$ satisfies single-crossing in the horizontal direction (which corresponds to changes in $v_1$ only), then there exists an optimal selling mechanism in which the first unit is sold deterministically.  We refer to a mechanism in which the first unit is sold deterministically as a {\sl line mechanism}. Line mechanisms are completely described by the payment for the first unit and the probability of allocating the second unit to types on the vertical line $(1,v_2)$, where $0\le v_2\le a$. If, in addition to horizontal single-crossing, $\Phi$ satisfies single-crossing in the vertical direction (which corresponds to changes in $v_2$ only) then there is an optimal mechanism which is {\sl semi-deterministic}, i.e., a line mechanism with at most one probabilistic value for allocating the second unit. Finally, if $\Phi$ satisfies diagonal single-crossing (along the diagonal boundary of the support of the distribution), in addition to horizontal and vertical single-crossing, then there exists an optimal mechanism that is deterministic.

Our results for increasing marginal values are under weaker conditions, in that horizontal single-crossing of $\Phi$ is sufficient for the existence of an optimal mechanism that is deterministic. In this optimal mechanism, the two units are bundled together and sold at a take-it-or-leave-it price.

We provide a class of distributions for decreasing marginal values, called the ordered decreasing values model, where our single-crossing conditions take a simple form and $\Phi$ is related to virtual utilities. Similarly, we introduce an ordered increasing values model with increasing marginal values.

To our knowledge, the function $\Phi$ is new to this literature. However, horizontal single-crossing of $\Phi$ is equivalent to a sufficient condition introduced by \cite{MM88}.  A version of $\Phi$ may be useful in proving the optimality of deterministic mechanisms in other settings, such as the sale of heterogeneous objects in more general models.

\medskip
\noindent
{\sc Related Literature:}  Early work on mechanism design with multidimensional types includes \citet{Ro87}, \cite{MM88}, \cite{Wi93}, \cite{Ar96}, and \cite{RC98}. As these papers focused primarily on divisible products, existence of deterministic mechanisms was not an issue.

\cite{Th04}, \cite{MV06}, \cite{MV07}, \cite{Py06}, \cite{Pa11a}, \cite{Pa11b}, and \cite{HR15} investigate the sale of indivisible, heterogeneous objects with independent, additive values. As already noted, it may be optimal to randomize in this setting. Moreover, as \cite{HR15} show, the optimal revenue may not be monotone in the distribution of the buyer's type.\footnote{We show that the sufficient conditions that imply the optimality of a deterministic mechanism in the decreasing and increasing marginal values models also imply revenue monotonicity.} Correlation between a buyer's values adds another layer of complexity and may increase the desirability of randomization. In a model with two heterogeneous goods and correlated values, \cite{HN19} show that mechanisms of bounded menu size, such as deterministic mechanisms, may yield a negligible fraction of the optimal revenue.

The optimality of deterministic mechanisms is investigated by \cite{MV06}, who obtains sufficient conditions in a model with two heterogeneous objects with independent, additive values; related papers include \cite{MHJ15} and \cite{TW17}. This question is also the focus of \cite{MV09} and \cite{DHP20} in a homogeneous objects model in which the buyer has the same privately known value for all units, but the number of units desired is privately known. In a general model, \cite{HH20} obtain sufficient conditions for the optimal mechanism to be bundling, which is a deterministic mechanism. In a model in which buyers purchase one of two heterogenous objects, \cite{Pa20} obtains an optimal mechanism (which may be random) and shows that if the optimal mechanism is deterministic, then it takes the form of selling the units separately.

When there are two or more buyers, \cite{CHLS19} provide sufficient conditions for the existence of an optimal Bayesian incentive compatible mechanism that is deterministic. These conditions do not apply to our setting, where there is one buyer, or to dominant strategy incentive compatible mechanisms. \cite{DDT17} and \cite{KM19} characterize optimality for a multi-product monopolist using duality theory.

There is a literature on approximately optimal mechanism design, starting with
the work of \citet{CHK07} and \citet{HR09}. Recent contributions include \citet{DRY15}, \citet{HN17}, \citet{HN19}, \citet{HR19}, \citet{BKKLRX20}, and \citet{BILW20}. These papers identify simple mechanisms, which are often deterministic mechanisms, that guarantee a constant fraction of the optimal mechanism revenue. These guarantees are usually independent of the prior. Another related  paper is \citet{Ca17}, which shows that posted-prices are robustly optimal for heterogeneous objects with additive values.

\medskip

The rest of the paper is organized as follows. We investigate the decreasing marginal values model   in Section~\ref{sec:dmv}. In Section~\ref{sec:det}, we provide a sufficient condition under which it is optimal to sell the first unit deterministically. Line mechanisms are characterized in Section~\ref{sec:lm} and sufficient conditions for the existence of an optimal mechanism that is deterministic are provided in Section~\ref{sec:odm}. Necessary conditions for a specific deterministic mechanism to be optimal are presented in Section~\ref{sec:nc}. {Two special cases of decreasing marginal values, the ordered decreasing values model and the conditional decreasing values model, are presented in Sections~\ref{sec:odvm} and \ref{sec:sdvm}. Our results for increasing marginal values are in Section~\ref{sec:imv}. In Section~\ref{sec:ext}, we describe the application of the results to heterogeneous objects and to a two-period model; we also show the optimal revenue is monotone under our sufficient conditions for deterministic optimality. All proofs are in an Appendix.

\section{Decreasing Marginal Values}\label{sec:dmv}

\aed{We present a model with decreasing marginal values (DMV) over two identical units of an indivisible object and prove optimality of a deterministic mechanism under single-crossing conditions.}

\medskip\noindent
\subsection{The DMV Model}

The buyer's value for the $i$th unit is $v_i$, $i=1,2$. The joint density function of $v=(v_1,v_2)$ is $f(v)$, which has support\footnote{All our results extend to type space $D=\{(v_1,v_2) \in [0,v^h] \times [0,av^h]: v_2 \le av_1\}$, where $v_h>0$. }
$$D\ \equiv\  \{(v_1,v_2)\in[0,1]\times[0,a] :  v_2\le a\,v_1 \}$$
Values are decreasing if $a\le 1$ (although we do not assume $a\le 1$ for any of our results). The density function $f(\cdot)$ is \rfr{continuously differentiable and strictly positive on its support.} As the support of the marginal distribution of $v_2$ depends on the realized value of $v_1$, the values $v_1$ and $v_2$ are not independent.\footnote{As is standard in screening literature, an equivalent interpretation of this model is that there is a continuum of buyers whose values over two units are decreasing and $f$ is the probability density function of this mass of buyers.}

An {\sl allocation rule} is a function $q=(q_1,q_2)$, where $q_i:D\to[0,1]$, $i=1,2$ is the (unconditional) probability that the $i$th unit is allocated to the buyer. If buyer type $(v_1,v_2)$ obtains a second unit, then this buyer must also obtain the first unit. Therefore, the feasibility of an allocation rule in this model is equivalent to the following constraint:
 $$q_1(v)\ge q_2(v),\qquad \forall v $$A {\sl transfer} is a function $t:D\to\Re$, a payment by the buyer to the seller. A {\sl mechanism} is~$(q,t)$.

The payoff of a buyer who truthfully reports $v$ is
\[
 u(v)\equiv v\cdot q(v) -t(v)
\]
A mechanism $(q,t)$ is {\sl individually rational} if $u(v)\ge 0$ for all $v$; it is {\sl incentive compatible} if
\[
 u(v)\ \ge \  u(v') + (v-v')\cdot q(v'),\qquad\forall v,v'
\]
It is well-known (see \cite{Ro87} or  \cite{Bo15}) that a necessary and sufficient condition for incentive compatibility is that $u(v)$ is a convex function and
\[
q_i(v) = \frac{\partial u(v)}{\partial v_i}, \qquad\mbox{a.e.}, \  i=1,2
\]
Thus
\[
t(v) = \grad u(v)\cdot v-u(v), \qquad\mbox{a.e.}
\]
The seller's expected revenue is
\begin{eqnarray}\label{eq:0}
\textsc{Rev}(q,t)\ \equiv \ \e[t(v)]& = & \int_{D} \Big[\grad u(v)\cdot v-u(v)\Big]f(v)dv
\end{eqnarray}
The integral of the first term in the integrand is the expected welfare from the mechanism. Subtracting the expected payoff of the buyer yields the seller's expected revenue.

A mechanism $(q^*,t^*)$ is  {\sl optimal} if it is incentive compatible (IC) and individually rational~(IR), and for any other IC and IR mechanism $(q,t)$ we have
$$\textsc{Rev}(q^*,t^*)\ \ge\ \textsc{Rev}(q,t)$$

It is easy to show that in any optimal mechanism $(q^*,t^*)$, if $q^*(v) = (0,0)$, then $t^*(v)=0$.
Thus, in an optimal mechanism, the payoff of a buyer type who received zero units is zero.

A mechanism $(q,t)$ is {\sl deterministic} if its allocation rule is deterministic, i.e., $q_i(v) \in \{0,1\}$ for all $v$ and $i$. If a mechanism is not deterministic, it is {\sl random}. A random mechanism (or allocation rule) is a lottery over deterministic mechanisms (or allocation rules).

Let $\cal Q$ be the set of IC and IR mechanisms. If mechanisms $(q^a,t^a)$ and $(q^b,t^b)$ are IC and IR then so is $\lambda (q^a,t^a)+(1-\lambda)(q^b,t^b)$, $\lambda\in[0,1]$. Thus, $\cal Q$ is a convex set. The set $\cal Q$ is compact.\footnote{See \cite{MV07} for a proof of compactness of the set of mechanisms for the sale of heterogeneous objects. A similar proof applies for the case of homogeneous objects considered in this paper.} As the expected revenue is a continuous, linear functional of $q$, it is maximized at an extreme point of $\cal Q$. When two or more indivisible objects are for sale, the extreme points of $\cal Q$ may be random mechanisms (see \cite{MV07}). This contrasts with the sale of one object to one buyer, where all extreme points of the set of IC, IR mechanisms are deterministic. Hence, a deterministic optimal mechanism always exists when a single object is sold to a buyer but a random mechanism might be optimal if two or more objects are sold.

\noindent
{\sl Note:} We say that a function $h$ is increasing if $x>x'$ implies $h(x)\ge h(x')$; it is strictly increasing if $x>x'$ implies $h(x)> h(x')$. We follow a similar convention for decreasing and strictly decreasing.

\subsection{The First Unit is Sold Deterministically}\label{sec:det}

In Proposition~\ref{prop:1} below, we show that under a sufficient condition on the density, there exists an optimal mechanism in which the first unit is sold deterministically. The following lemma is required for the proposition. The proof, which is in the Appendix, is similar to that of a result in \cite{MM88}; however, our assumption of decreasing marginal values yields a simpler expression for expected revenue.

\begin{lemma}\label{le:1} The seller's expected revenue from an IC and IR mechanism $(q,t)$ is
\begin{eqnarray*}
\textsc{Rev}(q,t) & = & \int\limits_0^au(1,v_2)f(1,v_2)dv_2 -\int\limits_0^a\int\limits_{\frac{v_2}a}^1 u(v_1,v_2)\Big[3f(v_1,v_2)+(v_1,v_2)\cdot\grad f(v_1,v_2)\Big]dv_1dv_2 
\end{eqnarray*}
\end{lemma}

The following condition on density, introduced by \cite{MM88}, is often invoked in the multidimensional mechanism design literature.

\medskip
\noindent
{\sl The density $f$ satisfies} {\bf Condition SC-H} {\sl if $3f(v)+v\cdot \grad f(v)\ge 0$ for almost all $v\in D$.}

\medskip

In the single-object case, Condition SC-H becomes $2f(v)+v\frac{df(v)}{dv}\ge 0$, which (i) is equivalent to the assumption that the expected revenue, $v[1-F(v)]$, is concave and (ii) implies that Myerson's virtual value function satisfies single crossing.\footnote{Condition SC-H is one of three conditions we impose on a function $\Phi$ that is defined in Section~\ref{sec:odm}; SC-H is a single-crossing assumption in the horizontal direction on $\Phi$. Together, the three conditions imply the existence of an optimal mechanism that is deterministic.} As shown next, under SC-H the first unit is allocated deterministically in an optimal mechanism.\footnote{If the inequality in SC-H is strict, then in any optimal mechanism $q_1(v)\in\{0,1\}$ for almost all $v$.}

\begin{proposition}\label{prop:1} If the density function $f$ satisfies Condition SC-H, then there exists an optimal mechanism $(q,t)$ in which $q_1(v)\in\{0,1\}$ for all $v$. That is, there exists an optimal mechanism such that for each $v$ \vspace{-3mm}
$$(q_1(v),q_2(v))\ \in\ \{ (0,0), (1,q_2) \} $$
where $q_2 \in [0,1]$ may depend on $v$. \aed{Moreover, $q_1(1,v_2)=1$ for all $v_2$ is optimal.}
\end{proposition}


\noindent
{\sc Remark 1:}  The proof of Proposition~\ref{prop:1} does not appeal to $q_1(v)\ge q_2(v)$, the feasibility constraint for the sale of identical objects. Even if it were feasible to have $q_1(v)<q_2(v)$, under SC-H there is an optimal mechanism in which $q_1(v)\ge q_2(v)$. Therefore, the proposition applies to heterogeneous objects as well. We elaborate on this in Section~\ref{sec:het}.


\medskip

Proposition~\ref{prop:1} is proved as follows. Lemma~\ref{le:1} implies that if, for any IC and IR mechanism,  the payoff function $u(v_1,v_2)$ is decreased when $v_1<1$ without decreasing $u(1,v_2)$ then, under Condition~SC-H, the expected revenue increases. One can make such decoupled changes to the buyer payoff function of a mechanism in which the first unit is allocated randomly to some buyer types, thereby creating a new IC and IR mechanism which has greater expected revenue. The argument is similar to the proof of Proposition~2 in \cite{Pa11b}, who showed that in the unit-demand case and in the additive, heterogeneous objects case, there is an optimal mechanism in which any positive allocation belongs to the upper boundary of the feasible allocation set.

Thus, under SC-H we may restrict our search for optimal mechanisms to those that allocate the first unit deterministically. This reduces the dimensionality of the problem as potentially optimal mechanisms are specified by a price for the first unit and an allocation rule for types $(1,v_2)$ only. We refer to such potentially optimal allocation mechanisms as line mechanisms.

\subsection{Line Mechanisms}\label{sec:lm}

WLOG, we restrict attention to {\it seller-favorable} mechanisms in the sense of \citet{HR15}. That is, when the buyer is indifferent between two or more outcomes, the buyer selects an outcome that maximizes the seller's revenue. A consequence is that $q_2(1,v_2)$ is right continuous in $v_2$.

Let $Y\equiv\{(1,v_2):v_2 \in [0,a]\}$ be the one-dimensional subset of the type space along the $v_2$-axis.

\begin{defn}
A mechanism $(q,t)$ is a {\bf line mechanism} if\vspace{-1mm}
\begin{enumerate}
\item[i.] its restriction to $Y$ is IC and IR
\item[ii.] for every $(1,v_2) \in Y$, $q_1(1,v_2)=1$
\item[iii.] for every $v \equiv (v_1,v_2) \in D \setminus Y$,
\begin{eqnarray*}
  \Big(q_1(v),q_2(v),t(v)\Big) &= \begin{cases}
  (0,0,0), & \textrm{if}~v_1 + v_2q_2(1,v_2) < t(1,v_2) \\
  (1,q_2(1,v_2),t(1,v_2)), & \textrm{otherwise}.
  \end{cases}
\end{eqnarray*}
\end{enumerate}
\end{defn}

\medskip
\noindent The first unit is allocated deterministically in a line mechanism, with types $(1,v_2)$ obtaining the first unit with probability one. A type $(v_1,v_2)$ is allocated $(q_1(1,v_2)=1,\,q_2(1,v_2),\,t(1,v_2))$ if it is IR; otherwise type $(v_1,v_2)$ gets $(0,0,0)$.  Thus,
  \begin{eqnarray}\nonumber
  u(v_1,v_2) &  = & \max\Big[0, v_1+v_2q_2(1,v_2)-t(1,v_2)\Big]\\
& = & \max\Big[0, u(1,v_2)-(1-v_1)\Big]\label{eq:revlm}
\end{eqnarray}


\begin{lemma}\label{le:2}
Every line mechanism is IC and IR on $D$.
\end{lemma}

The optimal mechanism in Proposition~\ref{prop:1} is a line mechanism. Hence, we  have the following corollary:
\begin{corollary}
\label{co:revdom}
If the density function $f$ satisfies Condition SC-H, then there is an optimal mechanism that is a line mechanism.
\end{corollary}

Corollary~\ref{co:revdom} simplifies the problem significantly, as a line mechanism is completely described by $t(1,0)$, the payment by type $(1,0)$ (i.e., the price for the first unit), and $q_2(1,v_2)$, the allocation rule for the second unit for types with $v_1=1$. However, the problem does not become one-dimensional. Two line mechanisms with the same allocation rule on $Y$ will have different allocation rules on $D$ if the price for the first unit is different in the two mechanisms. Moreover, the set of line mechanisms is not convex.

\bigskip\noindent
{\sc\bf The Structure of Line Mechanisms}\vspace{2mm}

\noindent
For any line mechanism $(q,t)$, define
\begin{eqnarray*}
Z_0(q,t):= \{(v_1,v_2):u(1,v_2)-(1-v_1) < 0\}
\end{eqnarray*}
Eq. (\ref{eq:revlm}) implies that the set of buyer types who do not receive any unit in the line mechanism is $Z_0(q,t)$. The closure of  $Z_0(q,t)$ consists of $(v_1,v_2)$ such that $u(v_1,v_2)=0$.  A line mechanism is shown in Figure \ref{fig:line1}.\footnote{Unless stated otherwise, in all figures
the $x$-axis denotes $v_1$ values and the $y$-axis denotes $v_2$ values.}

\begin{figure}[!hbt]
\centering
\includegraphics[height=2.3in]{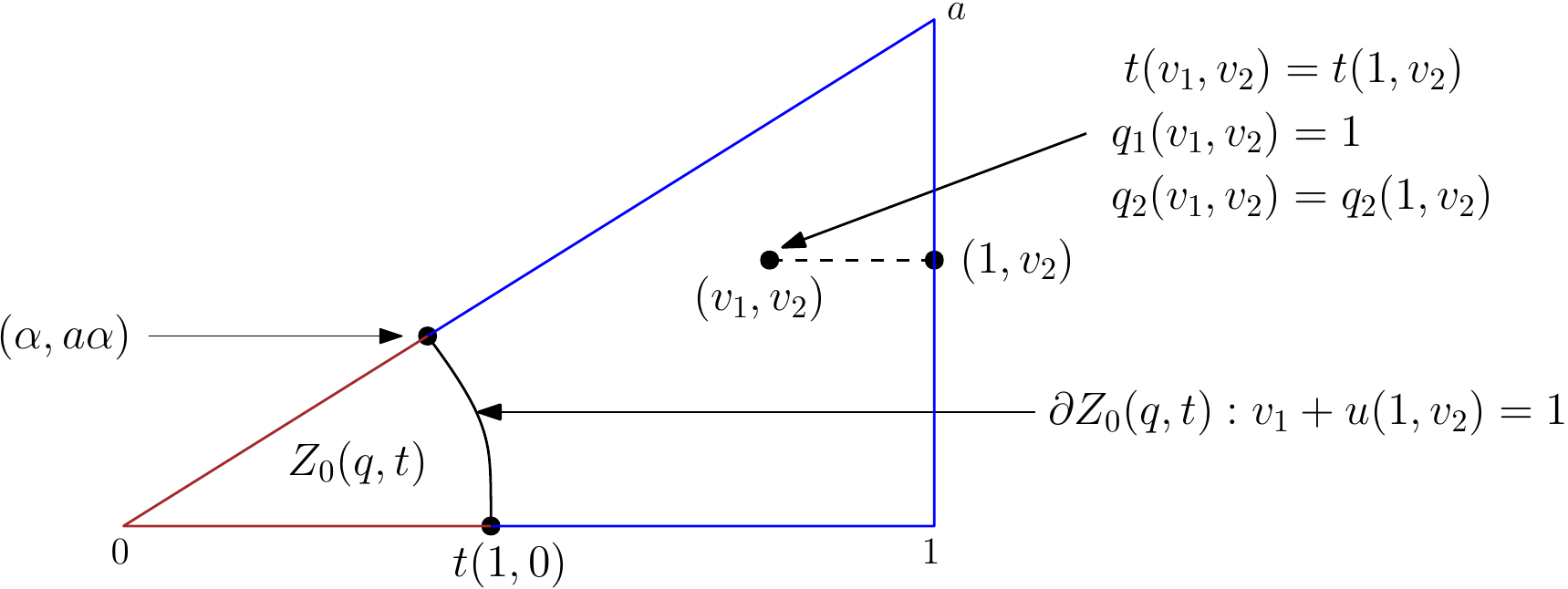}
\caption{A line mechanism}
\label{fig:line1}
\end{figure}

For any line mechanism $(q,t)$, define \aed{$\alpha\in[0,1]$ as the unique solution to}
\begin{equation}
\label{eq:alpha}
\alpha = 1-u(1,a\alpha)
\end{equation}
That $\alpha$ exists and is unique follows from \aed{$0\le 1-u(1,0)$, $1 \ge 1-u(1,a )$,} and the fact that $x+u(1,ax)$ is strictly increasing and continuous in $x$. The dependence of $\alpha$ on $(q,t)$ is suppressed in the notation.  From (\ref{eq:alpha}) we have
\begin{align}\label{eq:l8}
\alpha=1-u(1,a\alpha) \le 1-u(1,0) = t(1,0)
\end{align}

The upper boundary of $Z_0(q,t)$ is
$$\partial Z_0(q,t):= \{(v_1,v_2):u(1,v_2)-(1-v_1) =0\}$$
Note that  $\partial{Z}_0(q,t)$ is a curve with slope $-\frac1{q_2(1,v_2)}$ (see Lemma~\ref{le:3} in Appendix~\ref{prf:lm} for a proof) that connects the points $(t(1,0),0)$ and $(\alpha,a\alpha)$.  A deterministic mechanism is a line mechanism in which $\partial{Z}_0(q,t)$  is piecewise linear with (at most) two line segments, one vertical and the other with slope $-1$.

Lemma~\ref{le:41} below further limits the search for an optimal mechanism to a subset of line mechanisms defined next. For a line mechanism $(q,t)$, let
\begin{align*}
\bar{q}_2 :=
\begin{cases}
 \sup \limits_{v_2<a\alpha} \big[\,q_2(1,v_2)\,\big], & \mbox{if }\ \alpha>0\\
 0, & \mbox{if }\ \alpha=0
 \end{cases}
\end{align*}

\begin{defn}\label{d:clm}
A line mechanism $(q,t)$ is a {\bf constrained line mechanism} if  \\
\hglue 0.3in either (i) $q_2(1,a\alpha)=\bar q_2$ and for all $v_2 > a\alpha$,   $q_2(1,v_2) \in \{\bar{q}_2,1\}$ \\
\hglue 0.52in or (ii)  $q_2(1,a\alpha)=1$.
\end{defn}

\begin{figure}
  \centering
  \includegraphics[width=3.5in]{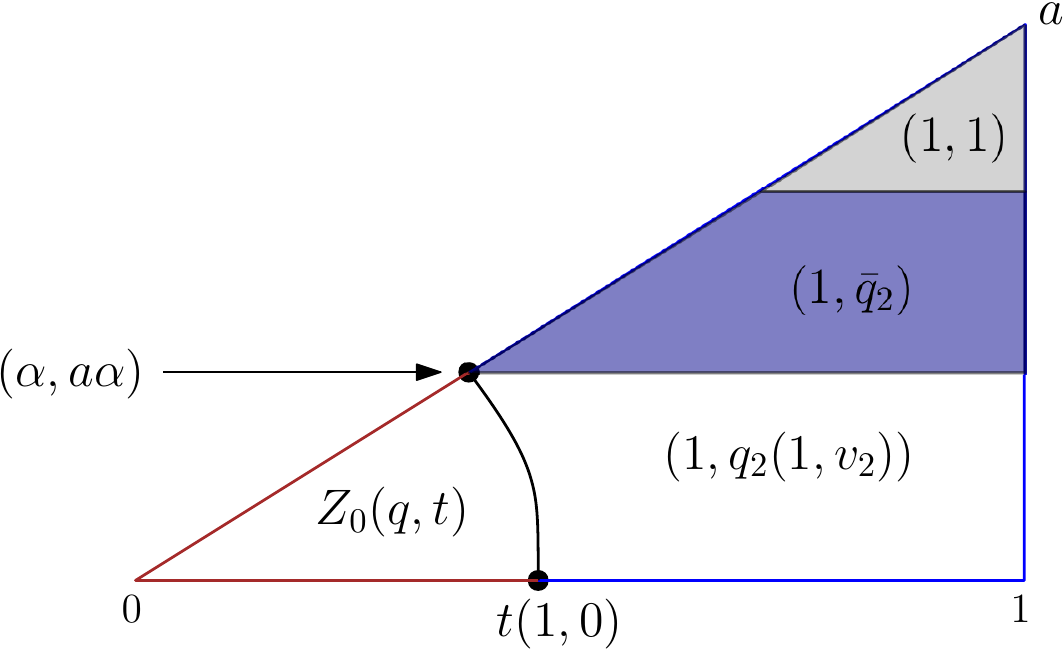}
  \caption{A constrained line mechanism}
  \label{fig:f4}
\end{figure}

If $\bar q_2=1$ then (i) and (ii) mean the same thing. Thus, in a constrained line mechanism for $v_2\ge a\alpha$ the probability of allocating the second unit takes at most two values, one of which may be less than 1 (see Figure~\ref{fig:f4}, where the case $q_2(1,a\alpha)=\bar q_2<1$ is illustrated). As noted earlier, $q_2(1,v_2)$ is right continuous WLOG. Therefore, if there is a discontinuity in $q_2(1,v_2)$ at $v_2=a\alpha$, then $q_2(1,a\alpha)=1$. However, for
$v_2<a\alpha$ any (increasing) value for $q_2(1,v_2)$ is possible.

\begin{lemma}\label{le:41}
If the density function $f$ satisfies Condition~SC-H, then there exists an optimal mechanism which is a constrained line mechanism.
\end{lemma}

Lemma~\ref{le:41} is proved by showing that there is an optimal mechanism which is an extreme point of a convex, compact subset of line mechanisms. As we are maximizing a linear function on this subset, the maximum is attained at an extreme point of this subset. Every extreme point of the subset is a constrained line mechanism.

\medskip\noindent
{\sc Remark 2:} From Definition~\ref{d:clm}, we conclude that if in a constrained line mechanism $\bar{q}_2=0$, then the mechanism is deterministic as $q_2(1,v_2)=0,\ \forall v_2<a\alpha$ and $q_2(1,v_2)\in\{0,1\},\  \forall v_2\ge a\alpha$. Therefore, in the sequel we restrict attention to constrained line mechanisms in which $\bar{q}_2>0$.  This, and the definition of $\bar{q}_2$, implies $\alpha>0$ and therefore (\ref{eq:l8}) implies $t(1,0) > 0$.

\subsection{Optimality of Deterministic Mechanisms}\label{sec:odm}

In this section, we provide sufficient conditions under which there is an optimal mechanism that is deterministic.

  It is useful to split the expected revenue of a constrained line mechanism $(q,t)$ into two parts,
  $$ \textsc{Rev}(q,t) = \textsc{Rev}^{\alpha-}(q,t)+\textsc{Rev}^{\alpha+}(q,t)$$
  where $\textsc{Rev}^{\alpha-}(q,t)$ is the expected revenue from types $v_2\le a\alpha$ and $\textsc{Rev}^{\alpha+}(q,t)$ is the expected revenue from types $v_2> a\alpha$. Define for every $(v_1,v_2)$,  \begin{eqnarray*}
  \Phi(v_1,v_2) &:= & f(1,v_2) - \int\limits_{v_1}^1\Big[3f(x,v_2) + (x,v_2)\cdot\nabla f(x,v_2)\Big]dx
  \end{eqnarray*}
\rfr{As $f$ is continuously differentiable, $\Phi$ is continuous.} Note that the function $\Phi$ depends only on $f$ and not on any mechanism. The role of  $\Phi$ is discussed after the next lemma. In Section~\ref{sec:odvm}, we show that in the ordered decreasing values model, which satisfies DMV, $\Phi$ is related to virtual utility.

  \begin{lemma}\label{lem:revline}
  If $(q,t)$ is a constrained line mechanism, then
    \begin{align*}
  \textsc{Rev}(q,t) &= \textsc{Rev}^{\alpha-}(q,t)+\textsc{Rev}^{\alpha+}(q,t),
  \end{align*}
  where
    \begin{align}
  \textsc{Rev}^{\alpha-}(q,t) &:= \int \limits_0^{a\alpha} \int \limits_{1-u(1,v_2)}^1 \Phi(v_1,v_2) dv_1 dv_2 \label{eq:rline1} \\
  \textsc{Rev}^{\alpha+}(q,t) 
 &:=  \int \limits_{a\alpha}^a \int \limits_{\frac{v_2}{a}}^1 \Phi(v_1,v_2) dv_1 dv_2
 + \int \limits_{a\alpha}^a u(\frac{v_2}{a},v_2)\Phi(\frac{v_2}{a},v_2)dv_2 \label{eq:rline2}
   \end{align}
  \end{lemma}

\medskip
As noted immediately after Proposition~\ref{prop:1}, decreasing $u(v_1,v_2)$ when $v_1<1$  and increasing $u(1,v_2)$ increases expected revenue (provided Condition SC-H is satisfied). These decoupled changes in $u$ are not possible in constrained line mechanisms. Whenever $u(1,v_2)$ is increased, $u(v_1,v_2)$ either increases or stays the same (see eq. (\ref{eq:revlm})). Thus, in a constrained line mechanism the net change in expected revenue due to an increase in $u(1,v_2)$, and the consequent increase in $u(v_1,v_2)$, may be positive or negative. This trade-off is captured by the function $\Phi$ (which we reiterate is independent of the mechanism and is only a function of the density).

To see the role of $\Phi$, consider a constrained line mechanism $(q,t)$ with buyer payoff $u(1,\cdot)$ for types in~$Y$. First, consider $v_2\le a\alpha$. Differentiating $\textsc{Rev}^{\alpha-}(q,t)$ with respect to $u(1,v_2)$, we see from (\ref{eq:rline1}) that  if $$\Phi(1-u(1,v_2),v_2)>0$$  then increasing $u(1,v_2)$ increases expected revenue. This is the process of ``straightening'' described later. If, instead,
$$\Phi(1-u(1,v_2),v_2)< 0$$
then decreasing $u(1,v_2)$ increases expected revenue. This is the process of ``covering'' a mechanism described later.  The single-crossing property  SC-V, introduced below, allows changes in $u(1,v_2)$ for a range of $v_2\le a\alpha$ in a manner that preserves incentive compatibility.

Similarly,  differentiating $\textsc{Rev}^{\alpha+}(q,t)$ with respect to $u(1,v_2)$ $(=u(\frac{v_2}{a},v_2)+ (1-\frac{v_2}{a})$) we see from (\ref{eq:rline2}) that for  $v_2>a\alpha$ if $u(1,v_2)$ is increased [decreased] when $\Phi(\frac{v_2}a,v_2)>0$ [$\Phi(\frac{v_2}a,v_2)<0$], then the expected revenue increases. The single-crossing property SC-D, introduced below, allows changes in $u(1,v_2)$  for a range of  $v_2> a\alpha$
in a manner that preserves incentive compatibility.

\medskip
Thus, $\Phi$ indicates the direction of revenue improvements, if any, for an arbitrary mechanism. Consider the following single-crossing properties of $\Phi$ in the horizontal, vertical, and diagonal directions in the type space:

\begin{defn}
The density function $f$ satisfies {\bf Condition SC} if
\begin{itemize}

\item[SC-H:] for every $v_2$, $\Phi$ is increasing in $v_1$

\item[SC-V:] for every $v_1$, $\Phi(v_1,\cdot)$ crosses zero at most once (from above). That is, for all $(v_1,v_2)$
\begin{eqnarray*}
\Big[ \Phi(v_1,v_2) > 0\Big] & \Longrightarrow & \Big[ \Phi(v_1,v'_2) > 0,~\forall~v'_2 < v_2\Big]
\end{eqnarray*}

\item[SC-D:] for every $v_2$, $\int_{v_2}^a \Phi(\frac{y}{a},y)dy$ crosses zero at most once (from below). That is, for all $v_2$
\begin{eqnarray*}
\Big[ \int \limits_{v_2}^a \Phi(\frac{y}{a},y)dy \ge 0 \Big] & \Longrightarrow & \Big[ \int \limits_{v'_2}^a \Phi(\frac{y}{a},y)dy \ge 0,~\forall~v'_2 > v_2\Big]
\end{eqnarray*}

\end{itemize}
\end{defn}

\noindent
Note that SC-H is equivalent to $3f(v)+v \cdot \nabla f(v) \ge 0$ for almost all $v$.  Further, SC-V is implied if $\Phi$ is decreasing in $v_2$ and SC-D is implied if $\Phi(\frac{y}a,y)$ satisfies single crossing.

\aed{ In Sections~\ref{sec:odvm} and \ref{sec:sdvm}, we provide two classes of distributions that satisfy Condition~SC.  As noted earlier, condition SC-H is commonly invoked in the literature. While SC-V and SC-D impose additional restrictions on the density $f$, we present an example in which if SC-H is satisfied then
so are SC-V and SC-D. Thus, in this example SC-V and SC-D are no more restrictive than SC-H.

\begin{example}\label{ex:sc}
{\em Let $a=1$ and
\begin{align*}
f(v_1,v_2) & = \frac{g(v_1)}{v_1},\qquad\forall 1\ge v_1\ge v_2\ge 0
\end{align*}
where $g$ is a density function with support $[0,1]$. In this example, the function $\Phi$ associated with $f$ is independent of $v_2$. Therefore, SC-V is satisfied for any density $g$. Moreover, if SC-H is satisfied then so is SC-D, but not vice versa.
See Appendix~\ref{prf:odm} for details. }\hfill$\Box$
\end{example}

Condition SC yields the main result in the DMV model.
}

\begin{theorem}
\label{theo:det}
If the density function $f$ satisfies Condition SC, then there is an optimal mechanism that is deterministic.
\end{theorem}

The proof consists of two  steps.\vspace{-3mm}
\begin{itemize}
\item {\sc Step 1.} SC-H implies that there is an optimal mechanism that is a constrained line mechanism (Lemma \ref{le:41}).  In a constrained line mechanism, $q_2(1,v_2)$ takes at most two values for $v_2\ge a \alpha$ but may take any number of values for $v_2<a\alpha$. Under SC-H and SC-V, Proposition \ref{prop:semi} in Section~\ref{sec:osdm} shows that there is an optimal  mechanism which is {\sl semi-deterministic}; that is, the optimal mechanism is a constrained line mechanism in which $q_2(1,v_2)$ takes  at most three values for $v_2\in[ 0,a]$,
and only one of these three values is strictly between 0 and 1.\footnote{In a deterministic mechanism, for any $v_2$, $q_2(1,v_2)$ is either 0 or 1.}\vspace{-2mm}
\item {\sc Step 2.} If SC holds, a deterministic line mechanism is optimal in the class of semi-deterministic line
mechanisms, completing the proof of Theorem~\ref{theo:det}.
\end{itemize}

Next, we explain Step 1 in some detail. The proof of Step 2 is in Appendix~\ref{prf:odm}.

\subsubsection{Optimality of Semi-deterministic Mechanisms}\label{sec:osdm}

For a constrained line mechanism $(q,t)$, define\footnote{The dependence of $\underline{v}_2$, $\bar{v}_2$ on $q_2$ is suppressed in the notation.}
\begin{align} \label{eq:vlbar}
\underline{v}_2 & := \inf \{v_2 \in [0,1]: q_2(1,v_2) > 0\}\\ \nonumber
\bar{v}_2 & :=\sup \{v_2 \in [0,1]:q_2(1,v_2)<1\}
\end{align}
As IC implies that $q_2(1,v_2)$ is increasing in $v_2$, we have $\underline{v}_2 \le \bar{v}_2$ with equality only if $(q,t)$ is a deterministic mechanism. By Remark~2, we may assume that $\bar{q}_2>0$. Therefore, for small positive $\epsilon$, $q_2(1,a\alpha-\epsilon)>0$ and $\underline{v}_2 < a\alpha$. 

Consider the following definition.
\begin{defn}
A constrained line mechanism $(q^s,t^s)$ {\bf straightens} another constrained line mechanism $(q,t)$ at $\underline{v}_2^s \in (\underline{v}_2,a\alpha]$ if
\begin{align*}
u^s(1,v_2) &=  u(1,\underline{v}_2^s),~\qquad~\forall~v_2 \le \underline{v}_2^s \\
u^s(1,v_2) &=  u(1,v_2),~\qquad~\forall~v_2 \ge \underline{v}_2^s
\end{align*}
where $u$ and $u^s$ are the payoff functions induced by $(q,t)$ and $(q^s,t^s)$, respectively.
\end{defn}

\aed{
\begin{figure}
\centering
\includegraphics[width=3.5in]{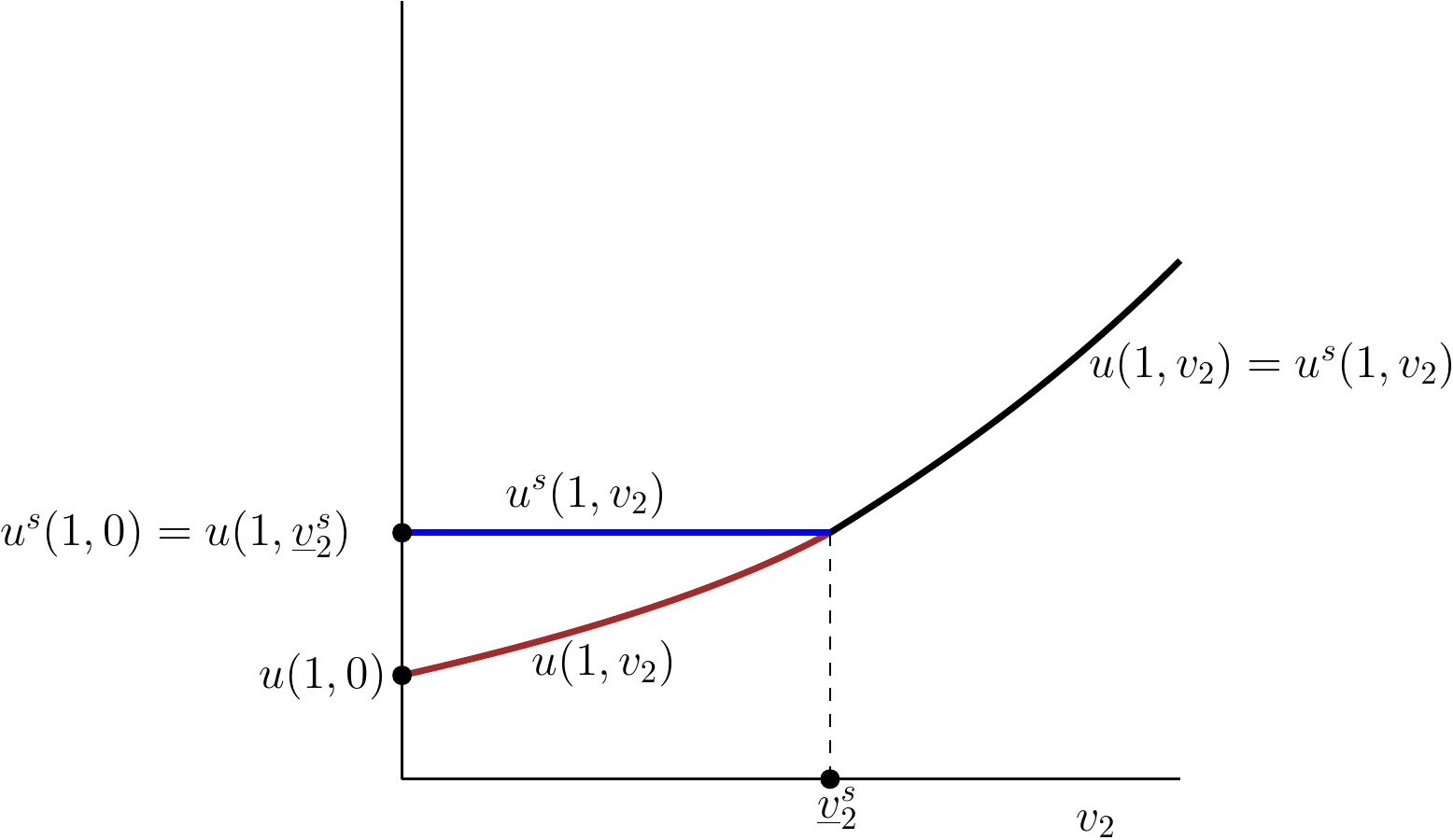}
\caption{Straightening generates convex $u^s$}
\label{fig:ust}
\end{figure}
Figure \ref{fig:ust} displays payoff functions $u^s(1,\cdot)$ and $u(1,\cdot)$ as a function of $v_2$.
As $(q,t)$ is IC and IR,  $u(1,v_2)$ is convex, increasing in $v_2$, and non-negative. Therefore, $u^s(1,v_2)=\max[u(1,v_2), u(1,\underline{v}_2^s]$ is also convex, increasing in $v_2$, and non-negative. Hence, $(q^s,t^s)$ restricted to $Y$ is IC and IR. As $(q^s,t^s)$ is a line mechanism,  Lemma~\ref{le:2} implies that it is IC and IR. }
Further, $(q^s,t^s)$ is completely specified by $u^s(1,\cdot\,)$ and
\begin{align*}
t^s(1,0) & = 1-u^s(1,0)=  1-u(1,\underline{v}_2^s)  <1-u(1,0) =t(1,0)\\
q_2^s(1,v_2) & =  \begin{cases}
0, & \mbox{if } v_2<\underline{v}_2^s\\
q_2(1,v_2), & \mbox{if }  v_2\ge\underline{v}_2^s
\end{cases}
\end{align*}
as illustrated in Figure \ref{fig:f6}. By construction, $\alpha=\alpha^s$ and, as $\underline{v}_2^s>\underline{v}_2$, we have $u^s(1,v_2) > u(1,v_2)$ for all $v_2 < \underline{v}_2^s$.
In a straightening, 
the price of the first unit is strictly  lower, $t^s(1,0)<t(1,0)$,  a buyer with $v_2<\underline{v}_2^s$ is never allocated a second unit, and the allocation of a buyer with $v_2\ge\underline{v}_2^s$ is unchanged. Consequently, $Z_0(q^s,t^s)\subsetneq Z_0(q,t)$.

\begin{figure}[!hbt]
\centering
\includegraphics[height=2.5in]{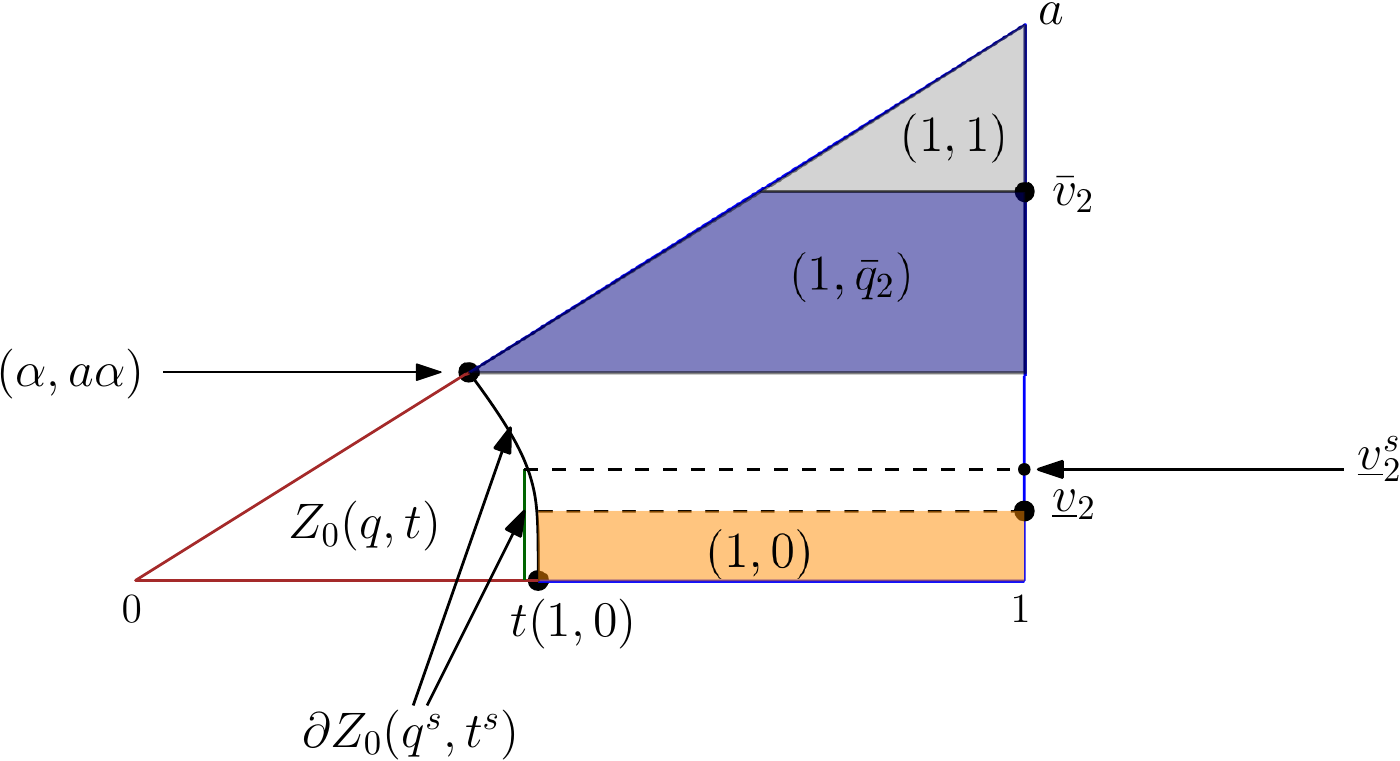}
\caption{Straightening a line mechanism}
\label{fig:f6}
\end{figure}

\begin{lemma}\label{lem:bound}
Suppose that the density function $f$ satisfies Conditions SC-H and SC-V.  Consider a constrained line mechanism $(q,t)$.  If
$$\Phi(t(1,0),\underline{v}_2) > 0,$$
then there exists a straightening $(q^s,t^s)$ of $(q,t)$, such that
$\textsc{Rev}(q^s,t^s) > \textsc{Rev}(q,t)$.
\end{lemma}

Next, consider the following two definitions.

\begin{defn}\label{def:sdm}
A constrained line mechanism $(q,t)$ is {\bf semi-deterministic} if
\begin{align*}
\Big(q_1(v),q_2(v)\Big) &\in \Big\{ (0,0),(1,0),(1,q_2 (1,a\alpha)),(1,1)\Big\},~\qquad~\forall~v \in D
\end{align*}
\end{defn}

In a semi-deterministic mechanism, $q_2(1,v_2)$ takes at most three values resulting in a  menu size of at most four. If $q_2 (1,a\alpha)= 0$ or 1 then the semi-deterministic mechanism is deterministic.

\begin{defn}\label{def:cov}
A mechanism $(q^c,t^c)$ is a {\bf cover} of a constrained line mechanism $(q,t)$ if
$q^c_1(1,v_2)=1$ for all $v_2 \in [0,a]$ and
\begin{align*}
\big(q^c_2(1,v_2),t^c(1,v_2)\big) =
\begin{cases}
\big(q_2(1,v_2),t(1,v_2)\big) & \textrm{if}~v_2 > a\alpha \\
\big(q_2(1,a\alpha),t(1,a\alpha)\big) & \textrm{if}~v_2 \le a\alpha~\textrm{and}~v_2q_2(1,a\alpha) \ge t(1,a\alpha)-t(1,0) \\
\big(0,t(1,0)\big) & \textrm{otherwise}
\end{cases}
\end{align*}

The extension of $(q^c,t^c)$ from $Y$ to $D$ is defined as in (\ref{eq:revlm}).
\end{defn}

Clearly, each constrained line mechanism has a unique cover.
\aed{
\begin{figure}
\centering
\includegraphics[width=3.5in]{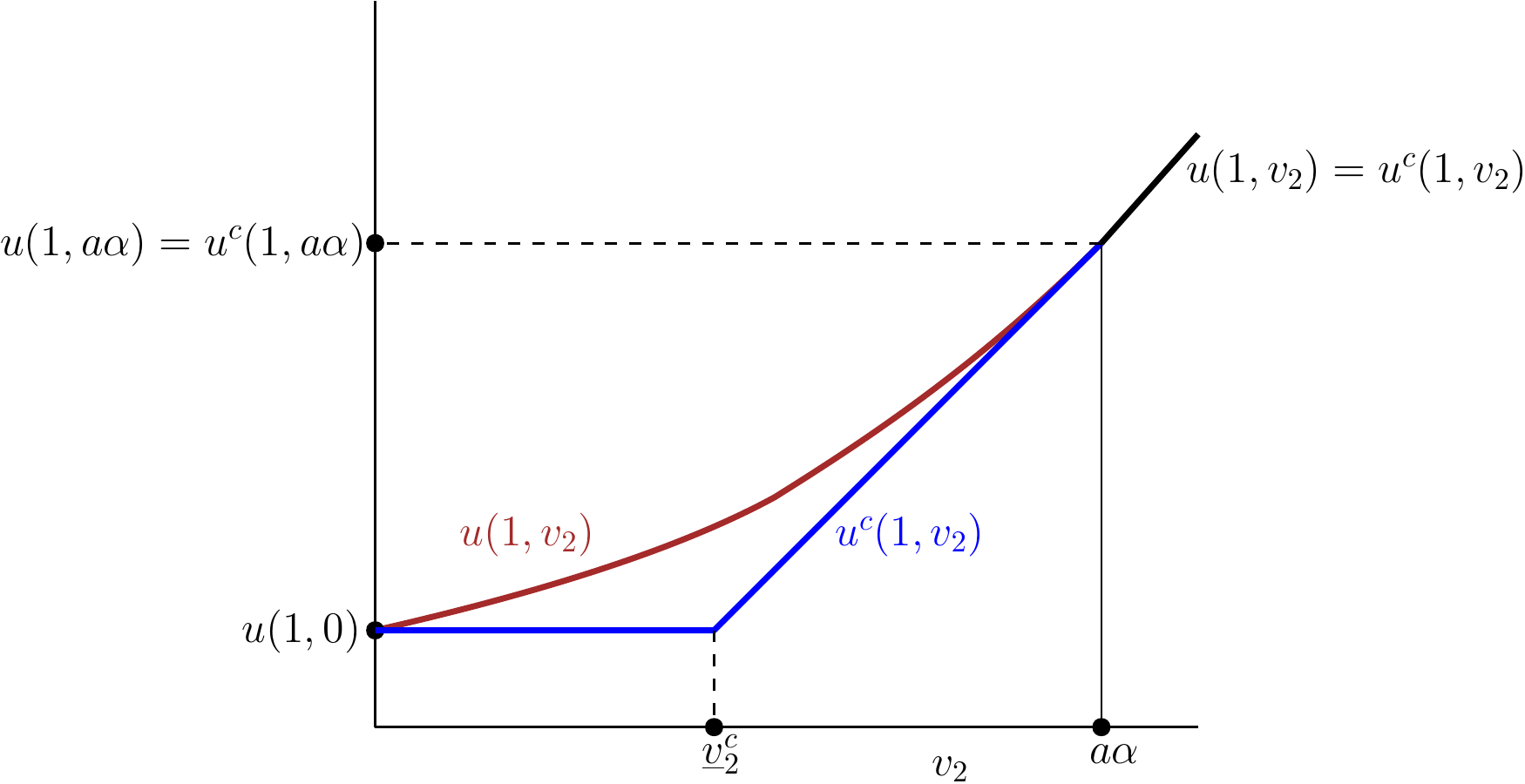}
\caption{Covering generates convex $u^c$}
\label{fig:cov}
\end{figure}
Let $u$ and $u^c$ be the payoffs induced by $(q,t)$ and $(q^c,t^c)$, respectively. Figure \ref{fig:cov} displays payoff functions $u(1,\cdot)$ and  $u^c(1,\cdot)$ as a function of $v_2$. Clearly, $u^c(1,\cdot)\ge u(1,0)\ge 0$. As the slope of $u^c(1,\cdot)$ is non-negative and increasing,  $u^c(1,\cdot)$ is convex and increasing in $v_2$. Hence, $(q^c,t^c)$ is IC and IR on $Y$. As $(q^c,t^c)$ is a line mechanism, Lemma~\ref{le:2} implies that it is IC and IR on $D$.}

The following properties of a cover are useful.

\begin{lemma}\label{le:covv}
Let $(q^c,t^c)$ be the cover of a constrained line mechanism $(q,t)$. Then
\begin{align} \label{eq:covdef1}
 q_2^c(1,a\alpha)&=q_2(1,a\alpha),  \quad  \alpha^c = \alpha\\ \nonumber
t^c(1,0) &= t(1,0) \\ \nonumber
u^c(1,v_2) &\le u(1,v_2),~\qquad~\forall~v_2 < a\alpha \\ \nonumber 
u^c(1,v_2) &=  u(1,v_2),~\qquad~\forall~v_2 \ge a\alpha
\end{align}
Further,  $(q^c,t^c)$ is semi-deterministic and if  $q_2^c(1,a\alpha)<1$, then $\bar{v}_2^c=\bar{v}_2>a\alpha$.
\end{lemma}

Figure~\ref{fig:newline3} shows the boundaries of type sets where 0, 1, $1+q_2(1,a\alpha)$ and 2 units are sold in the cover $(q^c,t^c)$ of a constrained line mechanism $(q,t)$.\footnote{\rfr{Figure~\ref{fig:newline3} is illustrative. One or more regions may be empty depending on $(q,t)$.}}
 Let $(q',t')$ be any semi-deterministic mechanism such that for any~$v$, $q_i(v)=0$ implies $q_i'(v)=0$. Then $Z_0(q^c,t^c)\subseteq Z_0(q',t')$.
 That is, among all semi-deterministic mechanisms that never allocate an object when $(q,t)$ does not, the cover of $(q,t)$ has the smallest set of types to which the object is never allocated.

\begin{figure}[!hbt]
\centering
\includegraphics[height=2.5in]{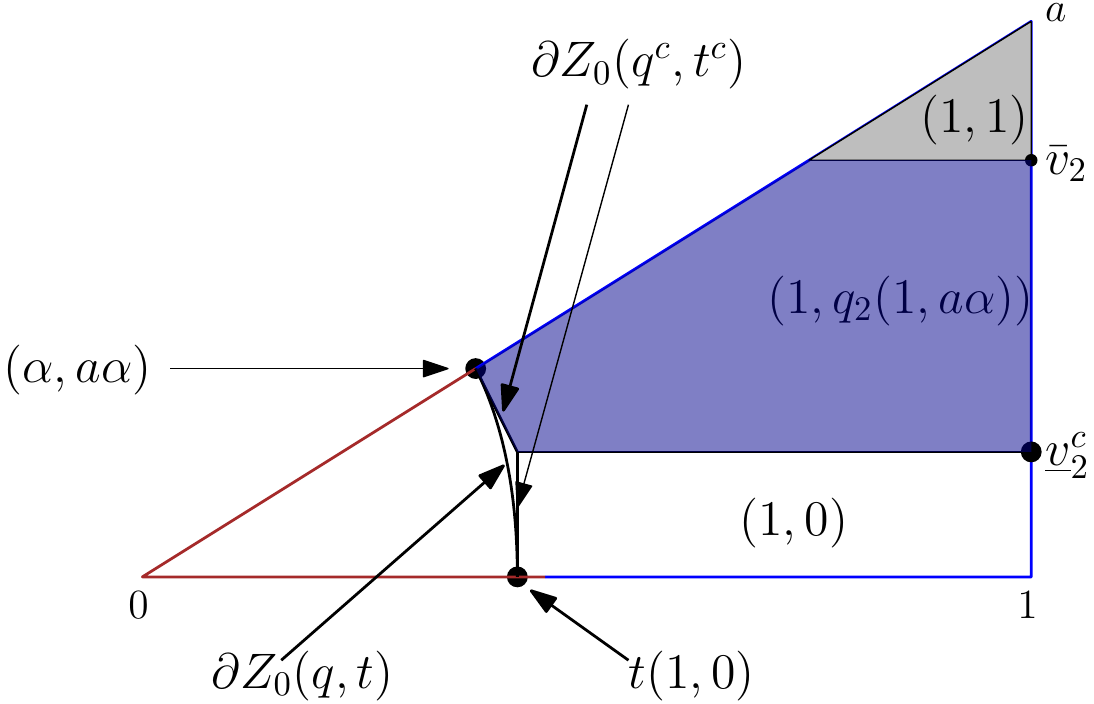}
\caption{Cover of a constrained line mechanism}
\label{fig:newline3}
\end{figure}

\bigskip

By Lemma~\ref{lem:bound}, $\Phi(t(1,0),\underline{v}_2) \le 0$ is a necessary condition for an optimal  constrained line mechanism $(q,t)$. Under this condition, we show that the revenue of any constrained line mechanism is no more than the revenue of its (semi-deterministic) cover.

\begin{lemma}\label{lem:cov1}
Suppose that the density function $f$ satisfies Conditions SC-H and SC-V. Consider a constrained line mechanism $(q,t)$. If
\begin{eqnarray}
\Phi(t(1,0),\underline{v}_2) &\le & 0 \label{eq:cov1},
\end{eqnarray}
then $\textsc{Rev}(q^c,t^c) \ge \textsc{Rev}(q,t)$ where $(q^c,t^c)$ is the cover of $(q,t)$.
\end{lemma}

\medskip
This leads to the main result of this section.

\begin{proposition}\label{prop:semi}
Suppose that the density function $f$ satisfies Conditions SC-H and SC-V. Then there exists an optimal mechanism that is semi-deterministic.
\end{proposition}

\noindent
{\bf Proof:}
Condition SC-H and Lemma~\ref{le:41} imply that there is an optimal mechanism $(q,t)$ which is a constrained line mechanism.
Therefore, Lemma~\ref{lem:bound} implies that (\ref{eq:cov1}) is satisfied for $(q,t)$.
Let $(q^c,t^c)$ be the (semi-deterministic) cover of $(q,t)$.  By Lemma \ref{lem:cov1}
\begin{eqnarray}\label{eq:deto}
\textsc{Rev}(q^c,t^c) & \ge & \textsc{Rev}(q,t)
\end{eqnarray}
Hence, $(q^c,t^c)$ is an optimal mechanism that is semi-deterministic.
\hfill$\blacksquare$

\medskip

Proposition~\ref{prop:semi} is used in the proof of Theorem~\ref{theo:det} in Appendix~\ref{prf:odm}\footnote{\rfr{In a model with two heterogeneous objects, \cite{TW17} obtain a similar result under the assumption that the elasticities of the densities of the (independently distributed) valuations are constant.}}

\subsection{Necessary Conditions for a Deterministic Optimal Mechanism}\label{sec:nc}

We provide necessary conditions for a deterministic mechanism to be optimal in the class of all deterministic mechanisms. If Condition SC is satisfied, then these conditions are necessary for optimality of a deterministic mechanism in the class of all mechanisms.

A deterministic mechanism is described by prices $p_i$, $i=1,2$ for the two units.
\rfr{So as to use first-order conditions, we restrict attention to cases when the optimal prices $(p_1^*,p_2^*)$ are in the interior of $D$, i.e., $p_1^*\in (0,1)$ and $p_2^*\in (0,a)$.\footnote{The assumption $p_1^*\le 1$ is without loss of generality because for any deterministic mechanism $(p_1,p_2)$ with $p_1>1$, the prices $\hat p_1=1$ and $\hat p_2=p_1+p_2-1$ yield the same expected revenue as $(p_1,p_2)$.}
Suppose that  $(p_1^*,p_2^*)$ satisfy $ap_1^* < p_2^*$. Then a necessary condition is that  $(p_1^*,p_2^*)$ are optimal in the set of all prices that satisfy $ap_1\le p_2$. That is,
\begin{eqnarray*}
(p_1^*,p_2^*) & \in & \arg\max_{(p_1,p_2)\in [0,1] \times[ap_1,a]}\ \Big[  p_1(1-F^1(p_1))+p_2(1-F^2(p_2)) \Big]
\end{eqnarray*}
where $F^i$ is the marginal cdf of $v_i$, $i=1,2$.  Differentiating with respect to $p_1$ and $p_2$ and equating to zero we have  $p_1^* =  \frac{1-F^1(p_1^*)}{f^1(p_1^*)}$, $p_2^* = \frac{1-F^2(p_2^*)}{f^2(p_2^*)}$, and $ap_1^* < p_2^*$ by assumption.


Next, assume that the (interior) optimal prices satisfy $ap_1^* > p_2^*$.\footnote{Note that $ap_1^* = p_2^*$ is not in the interior of $D$. Also, a bundling mechanism is not in the interior of $D$ since $p^*_2=0$ in a bundling mechanism.} The following necessary conditions are implied.

}

\begin{proposition}\label{prop:nec}
If \rfr{the optimal prices $(p_1^*,p_2^*)$ satisfy $p_1^*\in(0,1),\ p_2^*\in(0,ap_1^*)$,} then
\begin{eqnarray}
\int \limits_0^{p_2^*} \Phi(p_1^*,v_2) dv_2 &= & 0 \label{eq:nece1} \\
\int \limits_{p_2^*}^{a\alpha^*}\Phi((1+a)\alpha^*-v_2,v_2)dv_2 + \int_{a\alpha^*}^a\Phi(\frac{v_2}a,v_2)dv_2 &= & 0 \label{eq:nece2}
\end{eqnarray}
where $\alpha^*=\frac{p_1^*+p_2^*}{1+a}$. Further, $\Phi(p_1^*,p_2^*)\le 0$ and $\Phi(p_1^*,0)\ge 0$.
\end{proposition}

\subsection{Ordered Decreasing Values Model}\label{sec:odvm}

We describe a model in which values are based on the order statistics of two draws from the same distribution. Let $X_1,X_2$ be two i.i.d. random variables with cdf $G$ and density function $g$ that is strictly positive and differentiable on its support $[0,1]$. Let
$$v_1=\max\{X_1, X_2\}, \quad v_2=a\min\{X_1, X_2\}$$
Thus $av_1\ge v_2$.  We call this an {\bf ordered decreasing values} model. Note that
$$f(v_1,v_2) \ =\ \frac 2 ag(v_1)g(\frac{v_2}a), \qquad 1\ge v_1\ge \frac{v_2}a\ge 0$$
This model is a natural generalization of the maximum game of \cite{BK02} to two objects. In the interpretation of \cite{BBM20}  of the maximum game, the $X_i$'s represent the values from the different ways of using the object; the buyer will put the object to its best possible use. A similar interpretation applies to the ordered decreasing values model, where, if the buyer obtains one unit of the object, she will deploy it in its best usage and if she obtains two units, she will deploy them in the two best usages.

Another interpretation is that the buyer in the ordered decreasing values model is an intermediary who resells the units to two final consumers. The seller does not have access to the final consumers and can only sell the units to the intermediary. The final consumers have unit demand, their values are distributed i.i.d. and the realizations are known to the intermediary. If the intermediary purchases only one unit, she will resell it to the final consumer with a higher value.\footnote{These two interpretations assume that $a=1$.}

Let $\eta_g(x) :=\frac{x}{g(x)} \frac{dg(x)}{dx}$ be the elasticity of $g$. For every $(v_1,v_2)$, define
\begin{align}
W(v_1,v_2) &:= v_1 - \frac{1-G(v_1)}{g(v_1)} \Big[2+\eta_g(\frac{v_2}{a})\Big] \label{eq:wod} \\
W_{min}(v_2) &:= \frac{1}{a^2}\Big[v_2 - \frac{1-G_{min}(v_2)}{g_{min}(v_2)}\Big] \label{eq:wminod},
\end{align}
where $G_{min}(v_2)=1-[1-G(\frac{v_2}{a})]^2$ is the cumulative of the marginal probability distribution of $v_2$ and $g_{min}(v_2):=\frac{2}{a}g(\frac{v_2}{a})(1-G(\frac{v_2}{a}))$ is the marginal density of $v_2$. We can write $W$ and $W_{min}$ as follows:\footnote{See (\ref{eq:phi1}) and (\ref{eq:phi2}) in the proof of Proposition \ref{prop:orddec} in the Appendix.}
\begin{align*}
W(v_1,v_2) &= \frac{\Phi(v_1,v_2)}{\frac{2}{a} g(v_1)g(\frac{v_2}{a})} = \frac{\Phi(v_1,v_2)}{f(v_1,v_2)} \\
W_{min}(v_2) &= \frac{1}{ag_{min}(v_2)}\int \limits_{v_2}^a \Phi(\frac{y}{a},y)dy
\end{align*}
These equations and the functional form of $W$ and $W_{min}$ suggest that $\Phi$ admits a virtual utility representation. We make this precise in Proposition~\ref{prop:ordvu} later in this section.

Consider the following single-crossing conditions on $W$ and $W_{min}$.

\begin{defn}\label{d:scg}
For every $v_1$, $W(v_1,\cdot)$ {\bf crosses zero at most once (from above)} if for every $v_2$
\begin{align*}
\Big[ W(v_1,v_2) > 0\Big] \Longrightarrow \Big[ W(v_1,v'_2) > 0,~\forall~v'_2 < v_2\Big]
\end{align*}

$W_{min}$ {\bf crosses zero at most once (from below)} if for every $v_2$
\begin{align*}
\Big[ W_{min}(v_2) \ge 0 \Big] \Longrightarrow \Big[ W_{min}(v'_2) \ge 0,~\forall~v'_2 > v_2\Big]
\end{align*}
\end{defn}
Note that the second condition requires that the density $g_{min}$ satisfies \rfr{Myerson's regularity condition, i.e., $v_2 - \frac{1-G_{min}(v_2)}{g_{min}(v_2)}$ is increasing.} It is satisfied if $g$ has increasing hazard rate.

\medskip

The following proposition gives an equivalent condition for Condition SC in the ordered decreasing values model.

\begin{proposition}\label{prop:orddec}
In an ordered decreasing values model,\vspace{-3mm}
\begin{itemize}
\item[(i)] SC-H is satisfied if and only if $\eta_g(x)\ge -\frac  3 2, \ \forall x$;
\item[(ii)] SC-V is satisfied if and only if  $W(v_1,\cdot)$ crosses zero at most once for all $v_1$;
\item[(iii)] SC-D is satisfied if and only if $W_{min}$ crosses zero at most once.
\end{itemize}\vspace{-2mm}
Hence, in an ordered decreasing model, if (i), (ii), and (iii) are satisfied, then there is an optimal mechanism that is deterministic.
\end{proposition}

The conditions in Proposition \ref{prop:orddec} are easy to check. \rfr{The following densities, each of which has support $[0,1]$},  satisfy these conditions:
\begin{itemize}
\item the uniform family: $g(x)=\alpha x^{\alpha-1}$ with $\alpha \ge 1$;

\item a family of truncated exponential distribution: $g(x)=\frac{\lambda e^{\lambda x}}{e^{\lambda} - 1}$ with $\lambda > 0$;

\item a family of beta distributions: $g(x)=\frac{x^{\alpha-1}(1-x)^{\beta-1}}{\int_0^1x^{\alpha-1}(1-x)^{\beta-1}dx}$ with $\alpha \ge 1 \ge \beta$.

\end{itemize}

\rfr{In a model with two heterogenous objects and independently distributed values, \citet{MV06} provide
sufficient conditions for the optimal mechanism to be deterministic. These sufficient conditions include
SC-H and increasing elasticity of densities. Increasing elasticity implies SC-V in the
ordered decreasing values model, and hence, it is stronger than the necessary and sufficient condition for SC-V  in (ii) of Proposition \ref{prop:orddec}.}

\begin{example}\label{ex:unif} {\sc Uniform Distribution}\\
{\em
We describe the optimal mechanism for a uniform distribution on the domain $D$. The density~is
\begin{eqnarray*}
f(v_1,v_2) & = &
\begin{cases} \frac 2 a, & \mbox{if } 1\ge v_1\ge \frac{v_2}a\ge 0\\
							0, & \mbox{otherwise}
\end{cases}
\end{eqnarray*}

 \begin{figure}
      \centering
      \begin{subfigure}[b]{0.4\textwidth}
          \centering
          \includegraphics[width=4in]{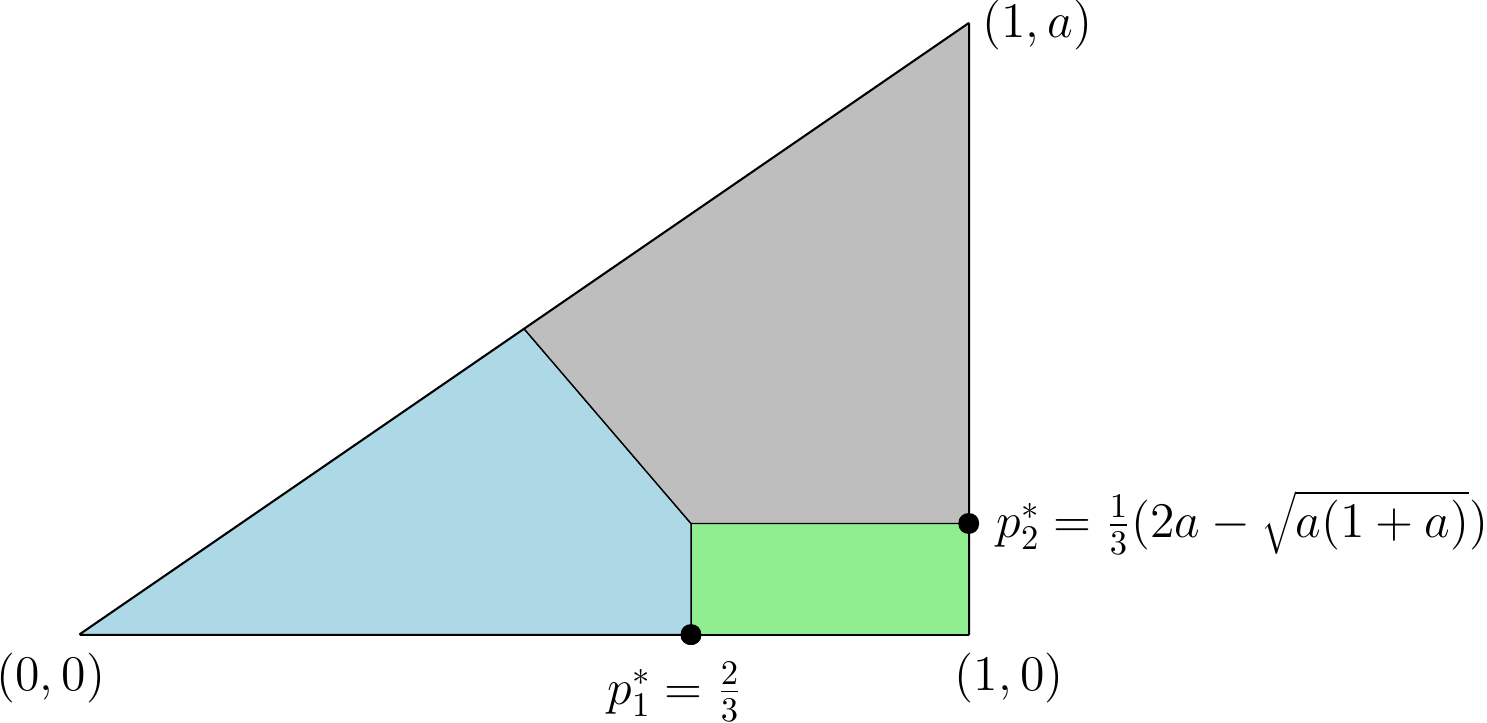}
          \caption{$a \ge \frac{1}{3}$.}
          \label{fig:un1}
      \end{subfigure}
      \hfill
      \begin{subfigure}[b]{0.4\textwidth}
          \centering
          \includegraphics[width=3in]{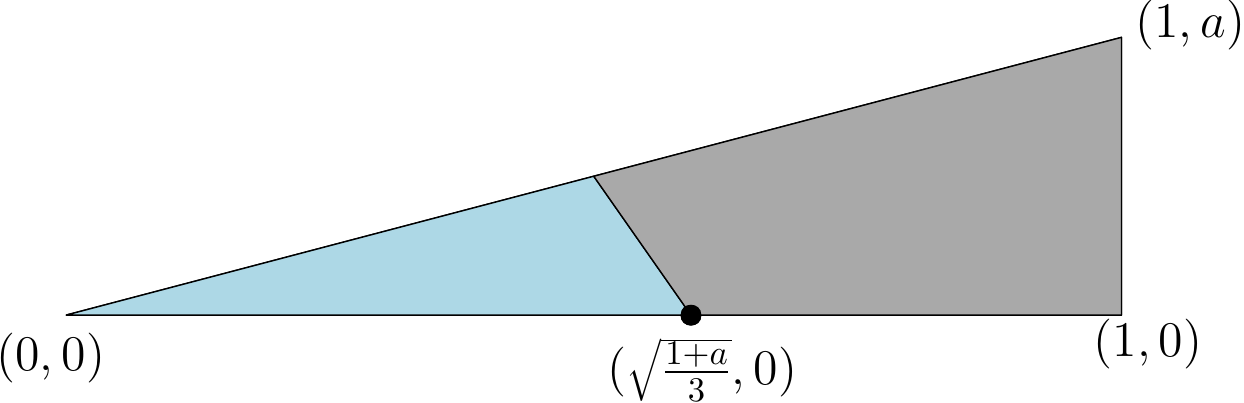}
          \caption{$a < \frac{1}{3}$.}
          \label{fig:un2}
      \end{subfigure}
      \caption{Optimal mechanism for uniform distribution.}
      \label{fig:unif}
 \end{figure}

For the uniform distribution,
it may be verified that in the class of deterministic mechanisms, it is optimal to set prices $(p^*_1,p^*_2)$ which satisfy $ap^*_1>p^*_2$. Moreover, $\Phi(v_1,v_2)= \frac{6v_1-4}a$. As $\Phi(2/3,v_2)=0$ for any value of $v_2$,  the only solution to eq. (\ref{eq:nece1}) is $p_1^*=2/3$. Therefore, the unique prices in the interior of $D$ that satisfy necessary conditions (\ref{eq:nece1}) and (\ref{eq:nece2}) for an internal optimal solution are\footnote{\citet{Ar16} shows that these are optimal prices for the case $a=1$.}
$$p_1^* = \frac{2}{3},\quad p_2^*=\frac 1 3 \Big(2a-\sqrt{a(1+a)}\,\Big)\qquad\qquad \mbox{\sc Unbundled prices}$$
A direct calculation reveals that the unique prices on the boundary of $D$ that are a candidate for an optimal solution are
$$p_1^* = \sqrt{\frac{1+a}{3}},\quad p_2^*=0 \qquad\qquad\qquad\qquad \mbox{\sc Bundle price}$$
As we are maximizing a continuous function on a compact set, an optimal solution exists. Therefore, one of these two prices is optimal. A calculation reveals that if $a>\frac 13$, then the optimal prices (i.e., optimal mechanism among all deterministic mechanisms) are the unbundled prices above. If, instead, $a<\frac 13$, then it is optimal to sell the two units as a bundle at the price $\sqrt{\frac{1+a}{3}}$.

In the limit as $a\to 0$, the buyer has positive value for one object only with density $f(v_1)=~2v_1$. The limit of the optimal bundling price as $a\to 0$ is $\sqrt{\frac{1}{3}}$, which is the optimal price for selling one object to a buyer with density $f(v_1)=2v_1$.

 The optimal prices are shown in Figure~\ref{fig:un1} for $a \ge \frac{1}{3}$ and in Figure \ref{fig:un2} for $a < \frac{1}{3}$.

That there is no random mechanism that yields greater expected revenue than these deterministic mechanisms follows from our results. First, note that the uniform model is an ordered decreasing values model with $v_1=\max\{X_1,X_2\}$ and $v_2=a\min\{X_1,X_2\}$, where $X_i$ are i.i.d. uniform on $[0,1]$. The uniform density on $[0,1]$ has elasticity 0 and has increasing hazard rate. Thus, the single-crossing conditions in Definition~\ref{d:scg} are satisfied. By Proposition~\ref{prop:orddec}, Condition SC is satisfied and by Theorem~\ref{theo:det} there is a deterministic mechanism that is optimal. \hfill$\Box$
}
\end{example}

As noted in the Introduction, \cite{HH20}  obtain sufficient conditions under which bundling is (or is not) optimal among all mechanisms. Applied to Example~\ref{ex:unif}, their sufficient conditions are as follows: (i) if the probability distribution of $\frac{v_1}{v_1+v_2}$ conditional $v_1+v_2=w$ is stochastically increasing\footnote{\rfr{A conditional distribution $\Pr[y|w]$ is stochastically increasing [decreasing] in $w$ if $\Pr[y|w]$ first-order stochastically dominates $\Pr[y|w'] $ when $w>w'$ [\,$w<w'\,]$.  }} in $w$, then bundling is optimal; (ii) if this distribution is strictly stochastically decreasing in $w$ then bundling is suboptimal.

A direct computation shows that for any $a>0$,
\begin{align*}
\Pr\bigg[\frac{v_1}{v_1+v_2}\le \frac 2{2+a} \,\bigg|\, v_1+v_2=w\bigg]& =
\begin{cases}
\frac1{2+a}, & \mbox{if } 0\le w\le 1 \\
\frac{aw}{(2+a)(1+a-w)}, & \mbox{if } 1\le w\le 1+ \frac a 2  \\
1, & \mbox{if  }  1+ \frac a 2 \le w
\end{cases}
\end{align*}
This distribution is stochastically decreasing in $w$, but it is not strictly stochastically decreasing. Thus, the sufficient conditions of Haghpanah and Hartline do not apply to Example~\ref{ex:unif}.

\subsubsection{Virtual Utility}\label{sec:vu}

In settings with one-dimensional types, the expected revenue of a selling mechanism equals the expected value under the mechanism of a virtual utility function of the buyer.
In the ordered decreasing values model, $W$ and $W_{min}$ are virtual utility functions. Proposition~\ref{prop:ordvu} below shows that the expected revenue of a deterministic mechanism equals the expectation of  $W$ and $W_{min}$ over regions of types defined below.

For any deterministic mechanism $(q,t) \equiv (p_1,p_2)$, define
\begin{align*}
S_1(p_1,p_2) &:= \{(v_1,v_2): q_1(v_1,v_2)=1\},~\  \qquad~\textrm{(types where one or two units are allocated)} \\
S^d_1(p_1,p_2) &:= \{(\frac{v_2}{a},v_2): q_1(\frac{v_2}{a},v_2)=1\},~\qquad~\textrm{(types on the diagonal where one unit is allocated)} \\
S^d_2(p_1,p_2) &:= \{(\frac{v_2}{a},v_2): q_2(\frac{v_2}{a},v_2)=1\},~\qquad~\textrm{(types on the diagonal where two units are allocated)}
\end{align*}
Note that if $ap_1 > p_2$, then $S^d_1(p_1,p_2) = \emptyset$. The expected revenue from $(p_1,p_2)$ is equal to a weighted expectation of $W$ and $W_{min}$ as shown next.

\begin{proposition}\label{prop:ordvu}
In the ordered decreasing values model, the expected revenue from a deterministic mechanism $(p_1,p_2)$ is
\begin{align}\nonumber
\textsc{Rev}(p_1,p_2) &= \e\Big[W(v_1,v_2)1_{\big\{ S_1(p_1,p_2)\big\}}\Big] + \e\Big[W_{min}(v_2) 1_{\big\{S^d_1(p_1,p_2) \big \}}\Big] \\ \label{eq:ordvu}
& \qquad +  \ (1+a)\e\Big[W_{min}(v_2) 1_{\big\{S^d_2(p_1,p_2) \big \}}\Big]
\end{align}
\end{proposition}

In (\ref{eq:ordvu}),  $1_{\{A\}}$ denotes the indicator function of event $A$.  The expectation of $W$ is taken over buyer types that buy at least one unit, while the expectation of $W_{min}$ is taken over the set of types along the diagonal that buy at least one unit.

 \cite{Ca17} and \cite{CDW16} obtained a virtual utility representation for multidimensional types.  Unlike their formulation, the virtual utility representation in Proposition~\ref{prop:ordvu} depends only on the primitives of the model and is not a function of the Lagrange multipliers of the revenue optimization problem.

\aed{

\subsection{Conditional Decreasing Values Models}\label{sec:sdvm}

Next, we present a model where the density $f$ is constructed using two densities $g_1$ and $g_2$ with support $[0,1]$ and cdfs $G_1$ and $G_2$ respectively.\footnote{We are grateful to the Associate Editor for suggesting this model. Although we assume that the support of $g_2$ is $[0,1]$,
the model can be modified so that the support of $g_2$ is $[0,a]$.}
The joint density $f$ is
\begin{align}\label{eq:sdvm}
f(v_1,v_2) & =  \frac{g_1(v_1)}{1-G_1(v_2)}g_2(v_2), \qquad 1\ge v_1\ge v_2\ge 0
\end{align}
To interpret this model, suppose one were to learn the value of $v_2$, drawn using marginal density $g_2$. Then, the value of $v_1$ is drawn from $[v_2,1]$ using conditional density $\frac{g_1(v_1)}{1-G_1(v_2)}$. We call this the {\bf conditional decreasing values} model.

Proposition~\ref{prop:sdvm2} gives sufficient conditions on $g_i$ so that $f$ satisfies SC.
These sufficient conditions are in terms of $\eta_{g_1}$ and $\eta_{g_2}$, the elasticities of densities $g_1$ and $g_2$.

\vglue 0.2in

\begin{proposition}\label{prop:sdvm2}
In a conditional decreasing values model, \vspace{-3mm}
\begin{itemize}
\item[(i)] SC-H is satisfied if $\eta_{g_1}(v_1)+\eta_{g_2}(v_2)\ge -3$, $\forall v_1\ge v_2$;
\item[(ii)] SC-V is satisfied if  $\eta_{g_2}(v_2) +v_2\frac{g_1(v_2)}{1-G_1(v_2)}$ is increasing and $\eta_{g_2}(v_2)\ge -2$, $\forall v_2$;
\item[(iii)] SC-D is satisfied if $\eta_{g_2}(v_2)\ge -2$, $\forall v_2$.
\end{itemize}\vspace{-2mm}
Hence, if (i), (ii), and (iii) are satisfied, then there is an optimal mechanism that is deterministic.
\end{proposition}

If $g_1$ has increasing hazard rate, then $\eta_{g_2}$ need not be increasing
to satisfy the requirement that $\eta_{g_2}(v_2) +v_2\frac{g_1(v_2)}{1-G_1(v_2)}$ is increasing. Note also that if $g_1=g_2=g$, then the sufficient conditions of Proposition~\ref{prop:sdvm2} are
satisfied if $\eta_{g}(x) \ge -\frac 32$ for all $x$ and $\eta_g(x) +x\frac{g(x)}{1-G(x)}$ is increasing.

We describe some families of densities that satisfy the sufficient conditions of Proposition~\ref{prop:sdvm2}. Let $\mathcal{G}$ denote the set of densities discussed immediately after Proposition \ref{prop:orddec} (they
include the uniform family, truncated exponential, and a family of Beta distributions).
The following $g_1$ and $g_2$ satisfy the sufficient conditions of Proposition \ref{prop:sdvm2}.
\begin{itemize}
\item Any $g_1,g_2 \in \mathcal{G}$.
\item Any $g_2 \in \mathcal{G}$ and $g_1$ has increasing hazard rate and satisfies $\eta_{g_1}(x) \ge -3$.
\item Any $g_1 \in \mathcal{G}$ and  $g_2$ satisfies the sufficient conditions in Proposition \ref{prop:sdvm2}(ii).
\end{itemize}

The ordered decreasing values model of Section~\ref{sec:odvm} is a special case of the conditional decreasing values model. If $g_1(v_1)=g(v_1)$ and $g_2(v_2)= 2g(v_2)[1-G(v_2]$ in (\ref{eq:sdvm}), we get an ordered decreasing values model with density $g$. As the joint density $f$ is based on a single density $g$ in the ordered decreasing values model, it leads to a sharper (iff) characterization of Condition SC in Proposition~\ref{prop:orddec}.

}

\section{Increasing Marginal Values}\label{sec:imv}

While decreasing marginal values is a common assumption, there are settings where values are increasing. For instance, if the buyer is unfamiliar with the product and incurs a learning cost before using it, the value for the second unit may be higher than the value for the first unit. Alternatively, if there is a fixed cost of production, then the model resembles increasing marginal values. As described in Section~\ref{sec:2perimv}, increasing marginal values is also satisfied in a model where the object is used over two periods.

\medskip

The departure from the earlier model is that the domain of $v$ is the following triangle
\begin{align*}
D:= \{(v_1,v_2) \in [0,a]\times[0,1]: v_1 \le av_2\}.
\end{align*}
The definitions of a mechanism and its properties remain as in Section~\ref{sec:dmv}. In particular, the constraint $q_1(v)\ge q_2(v),\ \forall v$ is imposed. The density function $f$ has support $D$ and is assumed to be absolutely continuous.

The proof of the next result is similar to that of Proposition~\ref{prop:1}.  However, unlike Proposition~\ref{prop:1}, the constraint $q_1(v)\ge q_2(v)$ is needed in the proof.

\begin{proposition}\label{prop:2}
If the density function $f$ satisfies Condition SC-H, then there exists an optimal mechanism in which $q_1(v)=q_2(v)$ for all $v$.
\end{proposition}

There is an optimal mechanism in which probability of selling each unit is the same.\footnote{If there are $n$ units, then the conclusion of Proposition~\ref{prop:2} generalizes to ``there exists an optimal mechanism in which $q_{n-1}(v)=q_n(v)$ for all $v$.''} In other words, the seller bundles the two units and sells them as one object. Hence, \cite{RZ83} and \cite{My81} imply the following main result of the IMV model:

\begin{theorem}\label{theo:det2}
If the density function $f$ satisfies Condition SC-H, then it is optimal to bundle the two units and sell them at a take-it-or-leave-it price.
\end{theorem}

Thus, a deterministic mechanism is optimal in this model under weaker conditions than in the case of decreasing marginal values. \aed{In the IMV model, if it is optimal to sell any unit(s) to a buyer type then, under SC-H, it is also optimal to sell the second unit with the largest feasible probability. This converts the sale of two units into the sale of one unit (the bundle). Consequently, SC-H is sufficient for optimality of a deterministic mechanism. In contrast, in the DMV model, SC-H implies that if it is optimal to sell any unit(s) then it is optimal to maximize the probability of selling the first unit. This does not convert the mechanism into the sale of a bundle and additional conditions (SC-V and SC-D) are needed for optimality of a deterministic mechanism. }

\rfr{Let $T$ be the cdf and $\tau$ the density of $w\equiv v_1+v_2$. The bundle price in the optimal mechanism of Theorem~\ref{theo:det2} is easily computed when $T$ is regular: it is optimal to sell the two units as a bundle at a price $B$ that solves $B=\frac{1-T(B)}{\tau(B)}$.}


\subsection{Ordered Increasing Values Model}

This is the counterpart of the order decreasing values model.
Let $X_1,X_2$ be two i.i.d. random variables with density $g(\cdot)$ that is strictly positive on its support $[0,1]$. Let $v_1=a \min\{X_1, X_2\}$ and $v_2=\max\{X_1, X_2\}$. Thus $v_1\le av_2$.

The next result applies Theorem~\ref{theo:det2} to this model.

\begin{proposition}\label{pr:6}
In the ordered increasing values model, if $g$, the density of $X_i$, satisfies $\eta_g(x) \ge -\frac{3}{2}$, then it is optimal to bundle the two units and sell them at a take-it-or-leave-it price. Further, if $g$ has increasing hazard rate, then the optimal bundle price $B$ solves
$B=\frac{1-T(B)}{\tau(B)}$.
\end{proposition}

We end this section with an example to show that the sufficient condition of \cite{HH20} under which bundling is optimal does not imply SC-H, the sufficient condition in Theorem~\ref{theo:det2}.

\begin{example}\label{ex:a}
{\em Let $(v_1,v_2)$ be distributed in the unit square with $v_1\le v_2$ and density function
\begin{align*}
f(v_1,v_2) & =
\begin{cases}
 \frac{12}{11} (2-v_1^2), & \mbox{if } 0\le v_1\le v_2\le 1\\
0, & \mbox{otherwise}
\end{cases}
\end{align*}
The density function $f$ satisfies SC-H as
\begin{align*}
3f(v)+v\cdot\nabla f& = \frac{36}{11}(2-v_1^2) + (v_1,v_2)\cdot (- \frac{24}{11}v_1, 0) \\
			& = \frac{72-60 v_1^2}{11} \\
			& \ge 0, \qquad\qquad \forall~v_1\in [0,1], \ v_2\ge v_1
\end{align*}

It may be verified that for $w\le 1$
$$\Pr\bigg[\frac{v_1}{v_1+v_2}\le 0.5 \,\bigg|\, v_1+v_2=w\bigg] = \frac{96-w^2}{8(24-w^2)}$$
which is increasing in $w$. Hence, the sufficient condition of \cite{HH20} for a bundling optimal (see Section~\ref{sec:odvm}) is not satisfied in this example. However, as SC-H is satisfied, we conclude from Theorem~\ref{theo:det2} that bundling is optimal.}
\hfill $\Box$
\end{example}

\section{Extensions}\label{sec:ext}

In this section, we discuss how our results can be applied to two related models. We also show that the optimal revenue is monotone in the probability distribution of values.

\subsection{Heterogeneous Objects and the DMV Model}\label{sec:het}

Our results for the DMV model (Propositions \ref{prop:1} \& \ref{prop:semi} and Theorem \ref{theo:det})
apply even if objects are not identical. Unlike with identical objects,
the constraint $q_1(v) \ge q_2(v)$ need not hold with heterogeneous objects.
But as noted in Remark 1, Proposition~\ref{prop:1} applies to heterogeneous objects as the constraint $q_1(v)\ge q_2(v)$ is not used in its proof. The analysis after Proposition~\ref{prop:1} is restricted to line mechanisms, which satisfy $q_1(v) \ge q_2(v)$. Therefore, consider a model with two {\sl heterogeneous} objects where the buyer values are distributed with a density function $f$ with support $\{(v_1,v_2) \in [0,1]\times[0,a]: v_2 \le av_1\}$.
Then, a constrained line mechanism is optimal if $f$ satisfies Condition SC-H, a semi-deterministic line mechanism is optimal if $f$ satisfies Conditions SC-H and SC-V, and a deterministic mechanism is optimal if $f$ satisfies Condition~SC.

In general, there exist IC and IR mechanisms with $q_2(v)>q_1(v)$ in the heterogeneous objects model described in the previous paragraph. We describe three exceptions to this rule. In the settings below, one of the two heterogeneous objects can be sold only after the other object is sold. Consequently, we have $q_1(v)\ge q_2(v)$ in all feasible mechanisms.

\vspace{-3mm}
\begin{itemize}
\item[a)] After purchasing a product, the buyer might be offered a related product or service.  For instance, after purchasing a new car the buyer might also purchase an extended warranty. The buyer has a positive value for the warranty only if she buys the car. Such {\sl add-on sales} fit our model.
\item[b)] A seller who offers two versions of a product, basic or premium. The buyer's value for the basic product is $v_1$ and for the premium product is $v_1+v_2$.
\item[c)] There are two time periods, 1 and 2. The product is available at two levels of quality, L and H.  Quality L lasts one period only, while quality H lasts two periods. The buyer and seller transact before period 1. A product of quality L is consumed in the first period only, which the buyer values at $v_1$. A product of  quality H is consumed in both periods, which the buyer values at $v_1+v_2$. Thus, the product is consumed in period 2 only if it is consumed in period 1. With $q_1(v)$ as the probability of consuming the product in period 1 (i.e., a sale of a product of quality L or H) and $q_2(v)$ as the probability of consuming the product in period 2 (i.e., a sale of a product of quality~H), we have (i) $q_1(v)\ge q_2(v)$ and (ii) the expected  value to buyer type $(v_1,v_2)$ from an allocation rule $q$ is $v_1q_1(v)+v_2q_2(v)$. It is natural that $v_1\ge v_2$ because of discounting.
\end{itemize}

\subsection{A Two-period Model with Increasing Marginal Values}\label{sec:2perimv}

Unlike in the DMV model, Theorem \ref{theo:det2} for the IMV model uses the feasibility constraint
$q_1(v) \ge q_2(v)$ in the proof. Hence, Theorem \ref{theo:det2} does not apply to heterogeneous objects models in general. However, as with the DMV model, the results of the IMV model also apply when there is an order in which two objects can be sold. In addition, the IMV results apply to the two-period model described next.

A durable product may be sold either at the beginning of the first period or at the beginning of the second period.\footnote{This model differs from scenario c) in Section~\ref{sec:het} in that the product is of one quality and it lasts two periods. The transaction may occur either in the first period or in the second period.}  If the product is sold in the first period, the buyer consumes it in both periods. If, instead, it is sold in the second period, then only second-period consumption is possible.

It is convenient to label time by the number of periods left, including the current period. Thus, the {\sl first period is period $2$} and the {\sl second period is period $1$}. If the buyer purchases the product in period 2, she  consumes it in both periods (period 2 and 1, in that order).

The buyer's values are $v_1$ for consumption in period $1$ (the second period) and $v_2$ for consumption in period 2 (the first period). Increasing marginal values, $v_1 \le v_2$, follows from discounting. The values $(v_1,v_2)$ are known to the buyer at the beginning (no dynamics).

An allocation rule $Q$ determines two things: $Q_1(v)$, the probability of selling the product  in period~1 (the second period), and $Q_2(v)$, the probability of selling the product in period~2 (the first period). A natural restriction is that $Q_1(v)+Q_2(v) \le 1$, as the buyer who buys in period 2 will not also buy later in period 1. The expected value to buyer type $(v_1,v_2)$ from this allocation rule is
$$(v_1+v_2)Q_2(v) + v_1 Q_1(v)= v_1 [Q_1(v)+Q_2(v)]+ v_2Q_2(v)$$
Let $q_1(v):=Q_1(v)+Q_2(v)$ and $q_2(v):=Q_2(v)$. Thus, $1\ge q_1(v)\ge q_2(v)\ge 0$. Note that $q_1(v)$ is the probability that the buyer consumes the product in the second period only and $q_2(v)$ is the probability that the buyer consumes the product in both periods.

\subsection{Monotonicity of Optimal Revenue}\label{sec:mono}

\cite{HR15}  show that the optimal revenue in multi-object auctions may decrease as the probability distribution of values increases. They also provide two sufficient conditions under which the optimal revenue is monotone, one of which is satisfied in our setting.\footnote{Although the model investigated by \cite{HR15} is one with heterogeneous objects, the argument in the proof of their Theorem~4 applies to homogenous objects as well.}

\begin{proposition}\label{prop:mono}
Suppose that either (i) or (ii) below holds:\vspace{-3mm}
\begin{itemize}
\item[(i)] In the DMV model, the density function $f$ satisfies condition SC.  \vspace{-2mm}
\item[(ii)] In the IMV model,  the density function $f$ satisfies condition SC-H.
\end{itemize}\vspace{-2mm}
In either case, let $\hat f$ be a density function that dominates $f$ by first-order stochastic dominance.\footnote{\rfr{A density function $\hat f$ dominates $f$ by first-order stochastic dominance if and only if $\e_{\hat f}[u(v_1,v_2)]\ge \e_f [u(v_1,v_2)]$ for every increasing function $u:D\to \Re$.}} Then the optimal revenue from $f$ is less than or equal to the optimal revenue from~$\hat f$.
\end{proposition}

\noindent
{\bf Proof:} If (i) or (ii) is satisfied then by Theorems~\ref{theo:det} or \ref{theo:det2} there exists an optimal mechanism that is deterministic. \rfr{A deterministic mechanism in an identical objects model is symmetric in the sense of \cite{HR15}.} 
Theorem~4 of \cite{HR15} implies that the optimal revenue from $f$ is no more than the optimal revenue from~$\hat f$.\footnote{Note that $\hat f$ need not satisfy condition SC or SC-H.}
\hfill$\blacksquare$

\section{Discussion}\label{sec:dis}

There are several directions we hope to explore in future work. An obvious one is generalizing the results to more than two units. Proposition~\ref{prop:1} generalizes to the sale of $n>2$ units as follows.  Suppose that there are $n$ units for sale with $ D=\{(v_1,v_2,\ldots,v_n)| 0\le v_1\le 1,\,0\le v_i\le a_iv_{i-1},\ i \ge 2\}$.  If the inequality in Condition~SC-H is changed to
$$(n+1)f(v)+v\cdot \grad f(v)\ge 0, \quad \mbox{ for almost all }v\in D,$$
then it is optimal to sell the first unit deterministically. A generalization of condition SC would be required to obtain a deterministic optimal mechanism.


The strategy of proofs developed in this paper may be useful in other models. Our preliminary investigations indicate that the approach used here can be adapted to some settings with heterogeneous objects.

\rfr{We have assumed that the lowest possible value of the buyer's valuation for either unit is zero. \cite{Pa11a} shows that with heterogeneous objects, a deterministic mechanism may be optimal when the lower end of the support of valuations is zero but need not be optimal when the lower end of the support has positive values. It is an open question whether similar examples exist with identical objects.}

\newpage

\appendix

\section{Appendix}

\subsection{Proofs of Section \ref{sec:det}}\label{prf:det}

{\bf Proof of Lemma~\ref{le:1}:} From (\ref{eq:0}), the seller's expected revenue is
\begin{eqnarray*}
\textsc{Rev}(q,t) & = & \int_D \bigg[\grad u(v)\cdot v-u(v)\bigg]f(v)dv\\
 & = & \int_0^a\int_{\frac{v_2}a}^1 \bigg[\grad u(v_1,v_2)\cdot (v_1,v_2)-u(v_1,v_2)\bigg]f(v_1,v_2)dv_1dv_2 \\
 & = & \int_0^1\int_0^{av_1} \bigg[\grad u(v_1,v_2)\cdot (v_1,v_2)-u(v_1,v_2)\bigg]f(v_1,v_2)dv_2dv_1
\end{eqnarray*}
Observe that
\begin{eqnarray*}
\int\limits_{\frac{v_2}a}^1\frac{\partial u(v)}{\partial v_1}v_1f(v)dv_1 & = & v_1 u(v)f(v)\bigg|_{\frac{v_2}a}^1-\int\limits_{\frac{v_2}a}^1 u(v)\Big[f(v)+v_1\frac{\partial f(v)}{\partial v_1}\Big]dv_1\\
        & = & u(1,v_2)f(1,v_2)- \frac{v_2}au(\frac{v_2}a,v_2)f(\frac{v_2}a,v_2)\\
        & & \qquad -\int\limits_{\frac{v_2}a}^1 u(v)\Big[f(v)+v_1\frac{\partial f(v)}{\partial v_1}\Big]dv_1\\[4pt]
\Longrightarrow\quad\int\limits_0^a\int\limits_{\frac{v_2}a}^1\frac{\partial u(v)}{\partial v_1}v_1f(v)dv_1dv_2
    & = & \int\limits_0^au(1,v_2)f(1,v_2)dv_2
     -\ \int\limits_0^a\frac{v_2}au(\frac{v_2}a,v_2)f(\frac{v_2}a,v_2)dv_2\\
        & & \qquad -\int\limits_0^a\int\limits_{\frac{v_2}a}^1 u(v)\Big[f(v)+v_1\frac{\partial f(v)}{\partial v_1}\Big]dv_1dv_2
 \end{eqnarray*}
 Similarly,
 \begin{eqnarray*}
 \int\limits_0^{av_1}\frac{\partial u(v)}{\partial v_2}v_2f(v)dv_2 & = & v_2 u(v)f(v)\bigg|_0^{av_1}-\int\limits_0^{av_1} u(v)\Big[f(v)+v_2\frac{\partial f(v)}{\partial v_2}\Big]dv_2\\
        & = & av_1u(v_1,av_1)f(v_1,av_1) -\int\limits_0^{av_1} u(v)\Big[f(v)+v_2\frac{\partial f(v)}{\partial v_2}\Big]dv_2
\end{eqnarray*}
\begin{eqnarray*}
\Longrightarrow\quad \int\limits_0^1 \int\limits_0^{av_1}\frac{\partial u(v)}{\partial v_2}v_2f(v)dv_2dv_1 & = &
\int\limits_0^1av_1u(v_1,av_1)f(v_1,av_1)dv_1\\[4pt]
 & & \quad-\ \int\limits_0^1\int\limits_0^{av_1} u(v)\Big[f(v)+v_2\frac{\partial f(v)}{\partial v_2}\Big]dv_2dv_1
\end{eqnarray*}
By a change of variable $v_2=av_1$, we have
\begin{eqnarray*}
\int\limits_0^1av_1u(v_1,av_1)f(v_1,av_1)dv_1 & = & \int\limits_0^a\frac{v_2}au(\frac{v_2}a,v_2)f(\frac{v_2}a,v_2)dv_2
\end{eqnarray*}
Thus,
\begin{eqnarray*}
\int\limits_D[\grad u(v)\cdot v]f(v)dv & = & \int\limits_0^au(1,v_2)f(1,v_2)dv_2
            -\int\limits_0^a\int\limits_{\frac{v_2}a}^1u(v)\Big[2f(v)+v\cdot\grad f(v)\Big]dv_2dv_1
\end{eqnarray*}
and
\begin{eqnarray*}
\textsc{Rev}(q,t)  & = & \int\limits_0^au(1,v_2)f(1,v_2)dv_2 -\int\limits_0^a\int\limits_{\frac{v_2}a}^1u(v)\Big[3f(v)+v\cdot\grad f(v)\Big]dv_1dv_2
\end{eqnarray*}
\hfill$\blacksquare$

\bigskip\noindent
{\bf Proof of Proposition~\ref{prop:1}:}  Condition SC-H and Lemma~\ref{le:1} imply that if $u$ is modified to $\hat u$ (while maintaining IC and IR) such that
\begin{eqnarray}\label{eq:util}
\hat u(1,v_2)\ge u(1,v_2), \quad \forall~v_2   \quad \mbox{and\ \ } \hat u(v_1,v_2)\le u(v_1,v_2), \quad \forall (v_1,v_2) \mbox{ s.t. } v_1<1
\end{eqnarray}
then $\textsc{Rev}(\hat q,\hat t)\ge \textsc{Rev}(q,t)$.

Let $(q,t)$ be any IC and IR mechanism. WLOG, assume that $q(0,0)=(0,0),\ t(0,0)=0$. Let $Y=\{(1,v_2):v_2 \le a\}$. Define\vspace{-3mm}
\begin{eqnarray*}
\hat q_1(1,v_2) & =& 1,\qquad \hat q_2(1,v_2)\ =\ q_2(1,v_2)\\
  \hat t(1,v_2) & = & t(1,v_2)+ (1-q_1(1,v_2))
\end{eqnarray*}
and $\hat q(0,0)=(0,0),\ \hat t(0,0)=0$. In the mechanism $(\hat q,\hat t)$, the probability of getting the first unit is increased to 1 for types $(1,v_2)$ and the payment increased so as to leave such types indifferent between $(q,t)$ and $(\hat q,\hat t)$. Extend $(\hat{q},\hat{t})$ from $Y\cup \{(0,0)\}$ to $v\in D\setminus [Y\cup \{(0,0)\}]$ as follows:
\begin{eqnarray}\label{eq:qhat}
  \Big(\hat{q}_1(v),\hat{q}_2(v),\hat{t}(v)\Big) &= \begin{cases}
  (0,0,0), & \textrm{if}~v_1 + v_2\hat{q}_2(1,v_2) < \hat{t}(1,v_2) \\
  (1,\hat{q}_2(1,v_2),\hat{t}(1,v_2)), & \textrm{otherwise}.
  \end{cases}
\end{eqnarray}
So, the range of $(\hat q,\hat t)$ is $\{(0,0,0)\}$ and the outcomes for types $(1,v_2)\in Y$. Clearly, $(\hat q,\hat t)$ is IR on $D\setminus Y$.

In the mechanism $(\hat q,\hat t)$, type $(1,v_2)$ obtains payoff equal to that in $(q,t)$ as
\begin{eqnarray*}
\hat u(1,v_2) & = & (1,v_2)\cdot \hat q(1,v_2) - \hat t(1,v_2) \\
                & = & (1,v_2)\cdot  q(1,v_2) +(1-q_1(1,v_2)) -  [t(1,v_2)+(1-q_1(1,v_2))] \\
                & = & u(1,v_2)
\end{eqnarray*}
Thus, $(\hat q,\hat t)$ is IR on $Y$. That $(\hat q,\hat t)$ is IC on $Y$ follows from
\begin{eqnarray*}
\hat{u}(1,v_2) - \hat{u}(1,v'_2) &=& u(1,v_2) - u(1,v'_2)\ \ge\ (v_2-v'_2)q_2(1,v'_2) \ = \ (v_2-v'_2)\hat{q}_2(1,v'_2)
\end{eqnarray*}
where the inequality follows from IC of $(q,t)$.

We use the fact that $(\hat q,\hat t)$ is IC on $Y$ to prove that $(\hat q,\hat t)$ is IC on $D\setminus Y$. Consider any type $ (v_1,v_2) \in D \setminus  Y$. The payoff to this type from outcome $(1,\hat{q}_2(1,v'_2),\hat{t}(1,v'_2))$ is
\begin{eqnarray*}
v_1 + v_2 \hat{q}_2 (1,v'_2) - \hat{t}(1,v'_2) &= & (v_1 - 1) + (v_2-v'_2)\hat{q}_2(1,v'_2) + 1 + v'_2 \hat{q}_2(1,v'_2) - \hat{t}(1,v'_2) \\
&= & (v_1 - 1) + (v_2-v'_2)\hat{q}_2(1,v'_2) + \hat{u}(1,v'_2) \\
&\le & (v_1 - 1) + \hat{u}(1,v_2) \\
& = & v_1 + v_2 \hat{q}_2(1,v_2)-\hat{t}(1,v_2),
\end{eqnarray*}
where the inequality follows since $(\hat{q},\hat{t})$ is IC for any $(1,v_2)\in Y$. But $v_1 + v_2 \hat{q}_2(1,v_2)-\hat{t}(1,v_2)$ is the
payoff of type $(v_1,v_2)$ from the outcome $(1,\hat{q}_2(1,v_2),\hat{t}(1,v_2))$.
Hence, the payoff of type $(v_1,v_2)$ is maximized at the outcome  $(1,\hat{q}_2(1,v_2),\hat{t}(1,v_2))$. The payoff from this outcome is $v_1 + v_2 q_2(1,v_2) - t(1,v_2)$.

To summarize, if $v_1+v_2 \hat q_2(1,v_2) < \hat t(1,v_2)$, then type $(v_1,v_2)$ strictly prefers $(0,0,0)$ to all other outcomes in the range of $(\hat q,\hat t)$; otherwise, this type's payoff is maximized at the outcome $(1,\hat q_2(1,v_2),$ $\hat t(1,v_2))$. From (\ref{eq:qhat}) we see that $(\hat q,\hat t)$ is IC on $D \setminus Y$.

Finally, the payoff of type $(v_1,v_2)\in D\setminus Y$ that is allocated $(1,\hat{q}_2(1,v_2),\hat{t}(1,v_2))$ in the mechanism $(\hat{q},\hat{t})$ is
\begin{align*}
\hat{u}(v_1,v_2) & = \hat{u}(1,v_2) -(1-v_1)\\
&=u(1,v_2) -(1-v_1) \\
& \le u(v_1,v_2) + (1-v_1)q_1(1,v_2) -(1-v_1) \\
				& = u(v_1,v_2) -(1-v_1)(1-q_1(1,v_2))  \\
				& \le u(v_1,v_2)
\end{align*}
where the first inequality follows from the IC of $(q,t)$ and the second from $v_1<1$. If, instead, $(q_1(v_1,v_2), q_2(v_1,v_2),t(v_1,v_2))=(0,0,0)$ then $\hat u(v_1,v_2)=0\le u(v_1,v_2)$ by IR of $(q,t)$.

Hence, $\hat{u}(1,v_2)=u(1,v_2)$ for all $(1,v_2) \in Y$ and $\hat{u}(v) \le u(v)$ for all $v \in D \setminus Y$. As the conditions in (\ref{eq:util}) are satisfied, we conclude that $\textsc{Rev}(\hat q,\hat t)\ge \textsc{Rev}(q,t)$. Therefore, as $(q,t)$ was arbitrary, there is an optimal mechanism in which the allocation of the first unit is deterministic.
\hfill$\blacksquare$

\subsection{Proofs of Section \ref{sec:lm}} \label{prf:lm}

{\bf Proof of Lemma~\ref{le:2}:} Fix a line mechanism $(q,t)$. By definition, $q_1(1,\cdot )=1$ and $(q,t)$ is IC and IR on $Y$. The rest of the proof is identical to the second part of the proof of Proposition~\ref{prop:1}. \hfill$\blacksquare$

\bigskip

The following lemma is needed in the sequel.

\begin{lemma} \label{le:3}
For any line mechanism $(q,t)$, the set $Z_0(q,t)$ satisfies the following properties.\vspace{-2mm}
\begin{enumerate}
\item[i.] $Z_0(q,t)$ is convex.
\item[ii.] Further, $\alpha \le t(1,0) \le 1$.
If $t(1,0)= \alpha $, then $q_2(1,y)=0$ for all $y \in [0,a\alpha)$.
\item[iii.] The slope of the boundary $\partial Z_0(q,t)$ is $-\frac 1{ q_2(1,v_2)}$.
\end{enumerate}
\end{lemma}

\medskip
\noindent
{\bf Proof:} \\ 
{\it i.} Take $v,v' \in Z_0(q,t)$ and let $v'' = \lambda v + (1-\lambda)v'$ for some $\lambda \in (0,1)$.
Then,
\begin{eqnarray*}
v''_1 + u(1,v''_2) &= & \lambda v_1 + (1-\lambda) v'_1 + u\big(1, \lambda v_2 + (1-\lambda)v'_2\big) \\
&\le & \lambda v_1 + (1-\lambda) v'_1 + \lambda u(1,v_2)+ (1-\lambda)u(1,v'_2) \\
&= &\lambda (v_1 + u(1,v_2)) + (1-\lambda) (v'_1+u(1,v'_2)) \\
 &< & 1
\end{eqnarray*}
where the first inequality follows from the fact that $u$ is convex and the second from the fact that $v,v' \in Z_0(q,t)$. Therefore, $v''\in Z_0(q,t)$.

%
%

\noindent {\it ii.} That  $\alpha \le t(1,0)$ follows from (\ref{eq:l8}) and $t(1,0)\le 1$ follows from IR  as $u(1,0)=1-t(1,0)\ge~0$.  If $\alpha=t(1,0)$, then $u(1,0)=u(1,a\alpha)=u(1,0)+\int_0^{a\alpha}q_2(1,y)$.
As $q_2$ is non-negative, we must have $q_2(1,y)=0$ for all $y \in [0,a\alpha)$.

\noindent {\it iii.} Differentiating along the boundary, $v_1 + u(1,v_2) = 1$, we get
\begin{eqnarray*}
1+\frac{\partial u(1,v_2)}{\partial v_2}\frac{d v_2}{d v_1} & = & 1+ q_2(1,v_2)\frac{d v_2}{d v_1} \ = \ 0 \\
\Longrightarrow\qquad \frac{d v_2}{d v_1} & = & -\frac 1{q_2(1,v_2)}
\end{eqnarray*}
\hfill$\blacksquare$

\bigskip
\noindent
{\bf Proof of Lemma \ref{le:41}:} We know that the buyer's payoff $u$ from any IC, IR mechanism $(q,t)$ satisfies $\nabla u = (q_1,q_2)$~a.e. WLOG we restrict attention to mechanisms with $u(0,0)=0$. Therefore,
$$u(v_1,v_2)\ =\ \int\limits_0^{v_1}q_1(s_1,0)ds_1+ \int\limits_0^{v_2}q_2(v_1,s_2)ds_2$$
Thus, $(\ref{eq:0})$ implies that the expected revenue functional is linear in the allocation rule $q$.

Let $(q^*,t^*)$ be a line mechanism that is optimal.  We know from Corollary~\ref{co:revdom} that such a mechanism exists. Let $\mathcal{Q}^{\alpha^*}$ be the set of line allocation rules that use $q_1^*$ for allocating the first unit and, for $v_2< a\alpha^*$, use $q_2^*$ for allocating the second unit. That is,
\begin{align*}
\mathcal{Q}^{\alpha^*}:= \{q': q'_1(v)=q^*_1(v)~\textrm{for all}~v~\textrm{and}~q_2'(v_1,v_2)=q_2^*(v_1,v_2)~\textrm{for all}~(v_1,v_2)~\textrm{such that}~v_2 < a\alpha^*\}
\end{align*}
Hence, for every line mechanism $(q',t')$ such that $q' \in \mathcal{Q}^{\alpha^*}$, we have
\begin{align*}
t'(1,0) & = t^*(1,0), \quad \alpha' = \alpha^*, \quad \mbox{and} \quad
u'(1,v_2) = u^*(1,v_2),\  \forall~v_2 \le a\alpha^*
\end{align*}
Recall that $\bar{q}_2 = \sup_{v_2 < a\alpha^*}\big[q_2^*(1,v_2)\big]$.
Let $\mathcal{Q}_2^{\alpha^*}$ be the set of all increasing functions $q'_2(1,\cdot)$ defined on $[a\alpha^*,a]$
such that $q'_2(1,a\alpha^*) \ge \bar{q}_2$ and $q'_2(1,a) \le 1$. Any $q'\in \mathcal{Q}^{\alpha^*}$ maps to a $q_2'\in \mathcal{Q}_2^{\alpha^*}$ and vice versa. Moreover, as \rfr{the line mechanism} $(q^*,t^*)$ maximizes expected revenue in the class of all IC and IR mechanisms, \rfr{the restriction of $q_2^*$ to domain $[a\alpha^*,a]$} must maximize expected revenue in $\mathcal{Q}_2^{\alpha^*}$.\footnote{That is, the mechanism corresponding to $q_2^*$ must maximize expected revenue in the subset of mechanisms corresponding to $\mathcal{Q}_2^{\alpha^*}$.}

The set $\mathcal{Q}_2^{\alpha^*}$ is convex as the convex combination of two increasing functions is increasing. Moreover, $\mathcal{Q}_2^{\alpha^*}$ is compact in the $L^1$-norm (see \cite{Bo15}, p. 16). As noted above, the expected revenue functional is linear in $q$ and therefore it is also linear in $q_2'\in \mathcal{Q}_2^{\alpha^*}$. Hence, the problem of maximizing expected revenue on the set $\mathcal{Q}_2^{\alpha^*}$ has a solution at an extreme point of $\mathcal{Q}_2^{\alpha^*}$. WLOG we may select the optimal $(q^*,t^*)$ to be such that $q_2^*$ is an extreme point of~$\mathcal{Q}_2^{\alpha^*}$.

We argue that every extreme point $q_2 \in \mathcal{Q}_2^{\alpha^*}$ satisfies $q_2(1,v_2) \in \{\bar{q}_2,1\}$ for all $v_2 \ge a\alpha^*$. Assume, instead,
that $q_2(1,v_2) \in (\bar{q}_2,1)$ for some $v_2 \ge a\alpha^*$. Define two line allocation rules $\hat{q}_2, \tilde{q}_2 \in \mathcal{Q}_2^{\alpha^*}$ as follows
\begin{eqnarray*}
\hat{q}_2(1,v_2) &= & \begin{cases}
2q_2(1,v_2) - \bar{q}_2, & \textrm{if}~\frac{1}{2}(1+\bar{q}_2) \ge q_2(1,v_2)\\
1, & \textrm{if}~q_2(1,v_2) > \frac{1}{2}(1+\bar{q}_2)
\end{cases}
\end{eqnarray*}
\begin{eqnarray*}
\tilde{q}_2(1,v_2) &= & \begin{cases}
\bar{q}_2, & \textrm{if}~\frac{1}{2}(1+\bar{q}_2) \ge q_2(1,v_2)\\
2q_2(1,v_2) - 1, & \textrm{if}~q_2(1,v_2) > \frac{1}{2}(1+\bar{q}_2)
\end{cases}
\end{eqnarray*}

\begin{figure}[!hbt]
\centering
\includegraphics[height=3in]{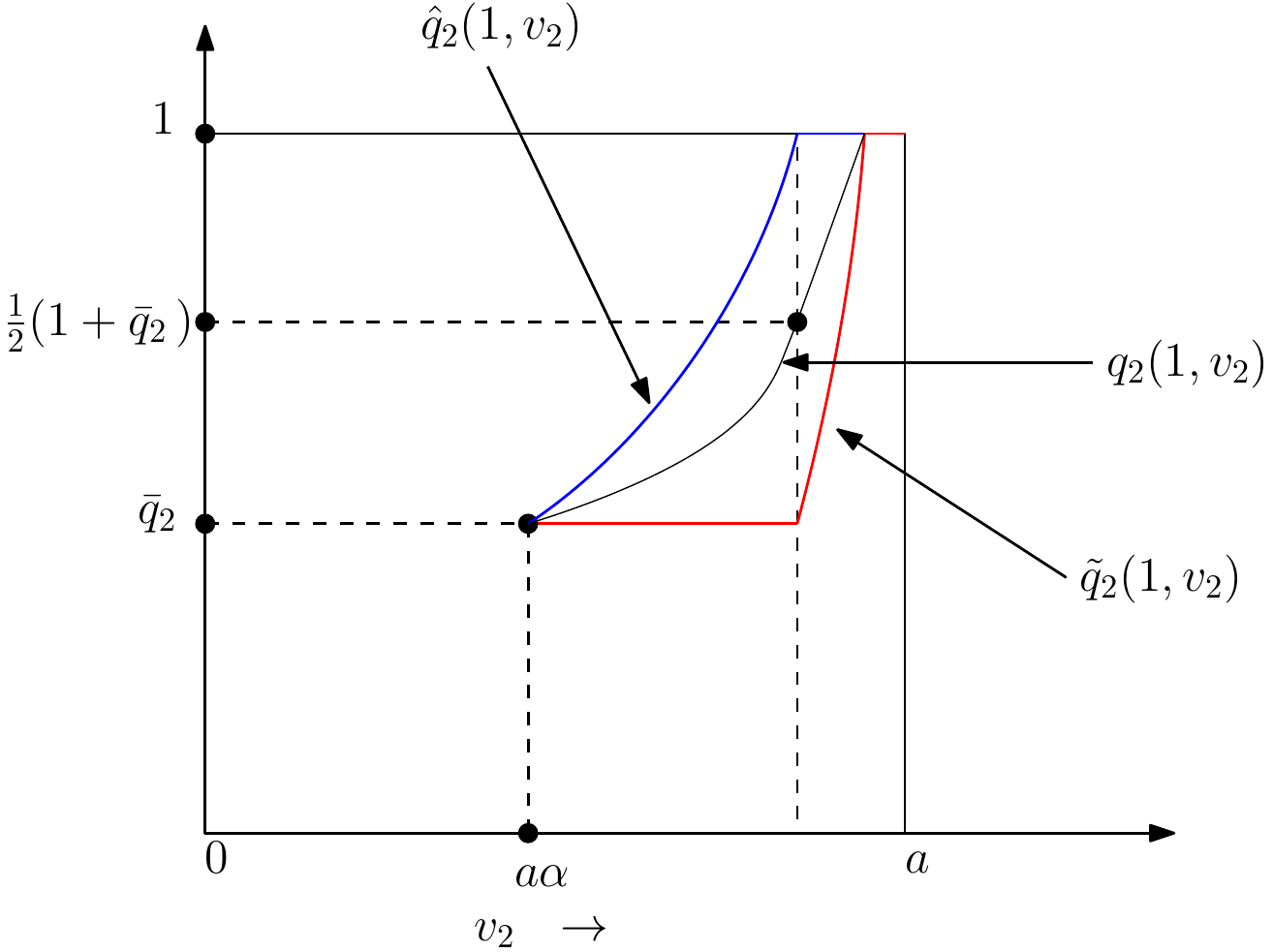}
\caption{Line allocation rules $\tilde{q}_2$ and $\hat{q}_2$}
\label{fig:line3}
\end{figure}

Both $\hat{q}_2(1,v_2)$ and $\tilde{q}_2(1,v_2)$ are increasing in $v_2$ and take values between $\bar{q}_2$ and 1 (see Figure \ref{fig:line3}).
Hence, $\hat{q}_2,\tilde{q}_2 \in \mathcal{Q}_2^{\alpha^*}$. As $q_2(1,v_2) \in (\bar{q}_2,1)$ for some $v_2 \ge a\alpha^*$,
we have $\hat{q}_2 \ne \tilde{q}_2 \ne q_2$. As $q_2=\frac{1}{2}[\hat{q}_2+\tilde{q}_2]$, $q_2$ cannot be an extreme
point of $\mathcal{Q}_2^{\alpha^*}$.

Note that if $\bar{q}_2=0$ then all extreme points of $\mathcal{Q}_2^{\alpha^*}$ are deterministic mechanisms.
\hfill$\blacksquare$

\subsection{Proofs of Section \ref{sec:odm}}\label{prf:odm}

\noindent
{\bf Proof of Lemma \ref{lem:revline}:}
By Lemma \ref{le:1},
  \begin{align*}
  \textsc{Rev}(q,t) &= \int \limits_0^a u(1,v_2)f(1,v_2)dv_2 + \int \limits_0^a \int \limits_{\frac{v_2}{a}}^1 u(v_1,v_2) \frac{\partial \psi(v_1,v_2)}{\partial v_1} dv_1 dv_2,
  \end{align*}
  where $\psi(v_1,v_2):=\int_{v_1}^1 \big[3f(x,v_2) + (x,v_2) \cdot \nabla f(x,v_2) \big]dx$.

  In a line mechanism,\footnote{Although the lemma is stated for constrained line mechanism, it is true for any line mechanism.} $u(v_1,v_2) = \max \big[0, v_1 - (1-u(1,v_2))\big]$. In particular,
  \begin{align*}
  u(v_1,v_2) =
  \begin{cases}
  0, & \textrm{if}~v_1 < \big(1-u(1,v_2)\big)~\textrm{and}~v_2 \le a\alpha \\
  v_1 - (1-u(1,v_2)), & \textrm{otherwise}
  \end{cases}
  \end{align*}
  As a result,
  \begin{align}\label{eq:3terms}
  \textsc{Rev}(q,t) &= \int \limits_0^a u(1,v_2)f(1,v_2)dv_2 \\ \nonumber
  &\quad + \int \limits_0^{a\alpha} \int \limits_{1-u(1,v_2)}^1 u(v_1,v_2) \frac{\partial \psi(v_1,v_2)}{\partial v_1} dv_1 dv_2 + \int \limits_{a\alpha}^a \int \limits_{\frac{v_2}{a}}^1 u(v_1,v_2) \frac{\partial \psi(v_1,v_2)}{\partial v_1} dv_1 dv_2
  \end{align}
We simplify each of the three terms in (\ref{eq:3terms}) below.\medskip

  \noindent   The first term can be written as
  \begin{align*}
  \int \limits_0^a u(1,v_2)f(1,v_2)dv_2 &= \int \limits_0^{a\alpha} \int \limits_{1-u(1,v_2)}^1 f(1,v_2)dv_1dv_2 +
  \int \limits_{a\alpha}^a u(1,v_2)f(1,v_2)dv_2 \\
  &= \int \limits_0^{a\alpha} \int \limits_{1-u(1,v_2)}^1 f(1,v_2)dv_1 dv_2 -
  \int \limits_{a\alpha}^a \Big(1-\frac{v_2}{a} - u(1,v_2)\Big)f(1,v_2)dv_2 \\
  &\quad + \int \limits_{a\alpha}^a (1-\frac{v_2}{a})f(1,v_2)dv_2 \\
  &= \int \limits_0^{a\alpha} \int \limits_{1-u(1,v_2)}^1 f(1,v_2)dv_1 dv_2 -
  \int \limits_{a\alpha}^a \Big(1-\frac{v_2}{a} - u(1,v_2)\Big)f(1,v_2)dv_2 \\
  &\quad + \int \limits_{a\alpha}^a \int \limits_{\frac{v_2}{a}}^1 f(1,v_2) dv_1 dv_2
  \end{align*}

  \noindent For the second term, use $u(v_1,v_2)=v_1 - (1-u(1,v_2))$ and $\psi(1,v_2)=0$ for all $v_2$ to obtain
  \begin{align*}
  &\int \limits_0^{a\alpha} \int \limits_{1-u(1,v_2)}^1 u(v_1,v_2) \frac{\partial \psi(v_1,v_2)}{\partial v_1} dv_1 dv_2 \\
  &= \int \limits_0^{a\alpha} \int \limits_{1-u(1,v_2)}^1 v_1 \frac{\partial \psi(v_1,v_2)}{\partial v_1} dv_1 dv_2 -
  \int \limits_0^{a\alpha} (1-u(1,v_2))\int \limits_{1-u(1,v_2)}^1 \frac{\partial \psi(v_1,v_2)}{\partial v_1} dv_1 dv_2 \\
  &=  \int \limits_0^{a\alpha} \Big[ v_1 \psi(v_1,v_2)\Big]_{1-u(1,v_2)}^1 dv_2 -  \int \limits_0^{a\alpha} \int \limits_{1-u(1,v_2)}^1 \psi(v_1,v_2)dv_1dv_2 + \int \limits_0^{a\alpha} (1-u(1,v_2))\psi(1-u(1,v_2),v_2)dv_2 \\
  &= -  \int \limits_0^{a\alpha} \int \limits_{1-u(1,v_2)}^1 \psi(v_1,v_2)dv_1dv_2
  \end{align*}
For the third term, again use $u(v_1,v_2)=v_1 - (1-u(1,v_2))$ to obtain
  \begin{align*}
  & \int \limits_{a\alpha}^a \int \limits_{\frac{v_2}{a}}^1 u(v_1,v_2) \frac{\partial \psi(v_1,v_2)}{\partial v_1} dv_1 dv_2 \\
  &= \int \limits_{a\alpha}^a \int \limits_{\frac{v_2}{a}}^1 v_1 \frac{\partial \psi(v_1,v_2)}{\partial v_1} dv_1 dv_2 -
  \int \limits_{a\alpha}^a (1-u(1,v_2)) \int \limits_{\frac{v_2}{a}}^1 \frac{\partial \psi(v_1,v_2)}{\partial v_1} dv_1 dv_2 \\
  &= \int \limits_{a\alpha}^a \Big[ v_1 \psi(v_1,v_2)\Big]_{\frac{v_2}{a}}^1 dv_2 -  \int \limits_{a\alpha}^a \int \limits_{\frac{v_2}{a}}^1 \psi(v_1,v_2)dv_1dv_2 + \int \limits_{a\alpha}^a (1-u(1,v_2)) \psi(\frac{v_2}{a},v_2)dv_2 \\
  &= \int \limits_{a\alpha}^a \Big(1- \frac{v_2}{a} - u(1,v_2)\Big) \psi(\frac{v_2}{a},v_2)dv_2 - \int \limits_{a\alpha}^a \int \limits_{\frac{v_2}{a}}^1 \psi(v_1,v_2)dv_1dv_2
  \end{align*}

 For all $(v_1,v_2)$, we have $\Phi(v_1,v_2)=f(1,v_2)-\psi(v_1,v_2)$. Therefore, inserting the three terms in (\ref{eq:3terms}), and noting that $u(\frac{v_2}a,v_2)=u(1,v_2)-(1-\frac{v_2}{a})$, we get
  \begin{align*}
  \textsc{Rev}(q,t) &= \int \limits_0^{a\alpha} \int \limits_{1-u(1,v_2)}^1 \Phi(v_1,v_2) dv_1 dv_2
  + \int \limits_{a\alpha}^a \int \limits_{\frac{v_2}{a}}^1 \Phi(v_1,v_2) dv_1 dv_2
  +\int \limits_{a\alpha}^a u(\frac{v_2}a,v_2)\Phi(\frac{v_2}{a},v_2)dv_2\\
  &= \quad \textsc{Rev}^{\alpha-}(q,t) \qquad\qquad\ \ +\qquad\qquad \qquad\textsc{Rev}^{\alpha+}(q,t)
  \end{align*}
  \hfill$\blacksquare$

\medskip
\noindent
{\bf Proof of Lemma \ref{lem:bound}:}
If two constrained line mechanisms $(q,t)$ and $(q',t')$ are identical for $v_2\ge a\alpha$ then $\alpha=\alpha'$, and $u(1,v_2)=u'(1,v_2)$, $\forall v_2\ge a\alpha$. It follows from (\ref{eq:rline2}) that
\begin{eqnarray}\label{eq:alphaplus}
\textsc{Rev}^{\alpha+}(q,t):=\textsc{Rev}^{\alpha'+}(q',t')
\end{eqnarray}

By assumption, $\Phi(1-u(1,0),\underline{v}_2) =\Phi(t(1,0),\underline{v}_2) > 0$.
The continuity of $\Phi$ and of $u$ implies that $\Phi(1-u(1,\underline{v}_2^s),\underline{v}_2^s) > 0$, where $\underline{v}_2^s:=\underline{v}_2+\epsilon$ and $\epsilon>0$ is small. As noted in Remark 2, we assume  ${\displaystyle \bar{q}_2=\sup_{v<a\alpha} [q_2(1,v_2)]>0}$ without loss of generality. Therefore, (\ref{eq:vlbar}) implies that $\underline{v}_2<a\alpha$ and we may take $\epsilon$ small enough such that $\underline{v}_2^s< a\alpha$.
By SC-V,
\begin{eqnarray}\label{eq:st1}
\Phi(1-u(1,\underline{v}_2^s),v_2)  &> & 0, \qquad \forall v_2\in[0,\underline{v}_2^s ]
\end{eqnarray}

Let $(q^s,t^s)$ be a straightening of $(q,t)$ at $\underline{v}_2^s$. Thus, $\alpha=\alpha^s$ and $u^s(1,v_2) \ge u(1,v_2)$, $\forall v_2 < \underline{v}_2^s$. Lemma~\ref{lem:revline}, eq. (\ref{eq:alphaplus}), and $u^s(1,v_2) = u(1,v_2)$, $\forall v_2 \ge \underline{v}_2^s$ imply
\begin{eqnarray*}
\textsc{Rev}(q^s,t^s) - \textsc{Rev}(q,t) &= & \textsc{Rev}^{\alpha-}(q^s,t^s) - \textsc{Rev}^{\alpha^s-}(q,t) \\
	& = & \int\limits_0^{\underline{v}_2^s} \ \int\limits_{1-u^s(1,v_2)}^{1-u(1,v_2)}\Phi(v_1,v_2)dv_1 dv_2 \\
	& = & \int\limits_0^{\underline{v}_2^s} \ \int\limits_{1-u(1,\underline{v}_2^s)}^{1-u(1,v_2)}\Phi(v_1,v_2)dv_1 dv_2 \\
&\ge& \int\limits_0^{\underline{v}_2^s} \ \int\limits_{1-u(1,\underline{v}_2^s)}^{1-u(1,v_2)}\Phi(1-u(1,\underline{v}_2^s),v_2)dv_1 dv_2 \\
&> &0,
\end{eqnarray*}
where the third equality follows from $u^s(1,v_2)=u(1,\underline{v}_2^s)$ for all $v_2 \le \underline{v}_2^s$, the first inequality  from SC-H, and the second inequality from (\ref{eq:st1}).
 \hfill$\blacksquare$

\noindent {\bf Proof of Lemma~\ref{le:covv}:}
By Definition~\ref{def:cov},
\begin{align}\label{eq:cov}
(q^c_1(1,v_2),q^c_2(1,v_2),t^c(1,v_2)) &:=
\begin{cases}
 \big(1,0,t(1,0)\big), & \textrm{if}~v_2 < \underline{v}_2^c \\
 \big(1,q_2(1,a\alpha),t(1,a\alpha)), & \textrm{if}~\underline{v}_2^c \le v_2 \le a\alpha
 \end{cases}
\end{align}
where $\underline{v}_2^c:=\frac{1}{q_2(1,a\alpha)}\Big[t(1,a\alpha) - t(1,0)\Big]\in[0,a\alpha]$. Thus, $ q_2^c(1,a\alpha)=q_2(1,a\alpha)$. Further, $u^c(1,a\alpha)=1+a\alpha q_2(1,a\alpha)-t(1,a\alpha)=u(1,a\alpha)=1-\alpha$ from (\ref{eq:alpha}). Hence, $\alpha^c=\alpha$.

If $\underline{v}_2^c>0$ then $t^c(1,0) = t(1,0)$. If, instead,  $\underline{v}_2^c=0$ then $t^c(1,0)=t(1,a\alpha) =t(1,0)$. Hence, $t^c(1,0) = t(1,0)$.

If $v_2 < \underline{v}_2^c$, then $u^c(1,v_2)=1-t(1,0)=u(1,0) \le u(1,v_2)$. If $a\alpha \ge v_2 \ge \underline{v}^c_2$, then $u^c(1,v_2)=1+v_2q_2(1,a\alpha)-t(1,a\alpha)= u(1,a\alpha) - (a\alpha-v_2)q_2(1,a\alpha) \le u(1,v_2)$, where the inequality follows from IC of $(q,t)$. Note that if $a\alpha=v_2$, we have $u^c(1,a\alpha)=u(1,a\alpha)$. Further, the definition of $(q^c,t^c)$ implies $u^c(1,v_2) =  u(1,v_2)$ for all $v_2 > a\alpha$.

As $u^c(1,v_2)\ge u(1,0)\ge 0$, $(q^c,t^c)$ is IR. Also, $u^c(1,v_2)$ is differentiable on $[0,a]\backslash \{\underline{v}_2^c,\bar{v}_2^c\}$; its derivative  is $0$ for $v_2 < \underline{v}^c_2$, $q_2(1,a\alpha)$ for $v_2 \in (\underline{v}_2^c,\bar{v}_2^c)$, and $1$ for $v_2 > \bar{v}_2^c$. Thus, $u^c(1,v_2)$ is convex and increasing in $v_2$. Hence, the line mechanism $(q^c,t^c)$ is IC and IR on $Y$ and, by Lemma~\ref{le:2}, it is IC and IR on $D$. Therefore, as $q^c_2(1,v_2)=q_2(1,v_2)$ for all $v_2 \ge a\alpha$, $(q^c,t^c)$ is a constrained line mechanism. Further, $q_2^c(1,v_2)\in\{0, q_2(1,a\alpha)\}$ for $v_2< a\alpha$. Hence, $(q^c,t^c)$ is semi-deterministic.

As $q_2(1,a\alpha)=q_2^c(1,a\alpha)<1$ and $(q,t)$ is right continuous, there exists an interval $[a\alpha,a\alpha+\delta)$ such that for every $v_2$ in this interval $q_2(1,v_2) =q_2(1,a\alpha) < 1$. Thus, $\bar{v}_2 > a\alpha$.  That $\bar{v}_2^c=\bar{v}_2$ follows from the fact $q_2^c(1,v_2)=q_2(1,v_2)$ when $v_2 > a \alpha$.
\hfill$\blacksquare$

\medskip

\bigskip
\noindent
{\bf Proof of Lemma \ref{lem:cov1}:}
Observe that (\ref{eq:cov1}) and SC-V imply that
\begin{eqnarray}\label{eq:cov2}
\Phi(t(1,0),v_2) &\le & 0,\qquad \forall v_2\ge \underline{v}_2
\end{eqnarray}
Let $(q^c,t^c)$ be the cover of $(q,t)$. By the covering property, $\alpha= \alpha^c$ and $u(1,v_2)=u^c(1,v_2)$, $\forall v_2\ge a\alpha$. Therefore, (\ref{eq:alphaplus}) implies $\textsc{Rev}^{ \alpha+}( q,  t)=\textsc{Rev}^{\alpha^c+}(q^c, t^c)$. This, together with Lemma~\ref{lem:revline}, implies
\begin{eqnarray*}
\textsc{Rev}(q,t)- \textsc{Rev}(q^c,t^c) &= & \textsc{Rev}^{\alpha-}(q,t) - \textsc{Rev}^{\alpha^c-}(q^c,t^c) \\
& = & \int \limits_{\underline{v}_2}^{a\alpha} \int \limits_{1-u(1,v_2)}^{1-u^c(1,v_2)}\Phi(v_1,v_2) dv_1 dv_2 \\
&\le &\int \limits_{\underline{v}_2}^{a\alpha} \int \limits_{1-u(1,v_2)}^{1-u^c(1,v_2)}\Phi(1-u^c(1,v_2),v_2)dv_1 dv_2 \\
&= &\int\limits_{\underline{v}_2}^{a\alpha}\big[u(1,v_2)-u^c(1,v_2)\big]\Phi(1-u^c(1,v_2),v_2) dv_2 \\
&\le&\int\limits_{\underline{v}_2}^{a\alpha} \big[u(1,v_2)-u^c(1,v_2)\big]\Phi(1-u^c(1,0),v_2) dv_2 \\
&= &\int\limits_{\underline{v}_2}^{a\alpha} \big[u(1,v_2)-u^c(1,v_2)\big]\Phi(t(1,0),v_2) dv_2 \\
&\le &0,
\end{eqnarray*}
where the first inequality follows from SC-H, the second inequality from $u(1,v_2)\ge u^c(1,v_2)$, $ \forall v_2<a\alpha$,  SC-H and $1-u^c(1,v_2) \le 1- u^c(1,0)$, $\forall v_2$, and the third inequality from $u(1,v_2)\ge u^c(1,v_2),\ \forall v_2<a\alpha$ and (\ref{eq:cov2}).  \hfill$\blacksquare$

\vspace{0.1in}

\noindent \aed{{\bf Calculations for Example~\ref{ex:sc}:}
 We have
\begin{align*}
3f(v) + v \cdot \nabla f(v) & = 3\frac{g(v_1)}{v_1}+v_1\frac{v_1g'(v_1)-g(v_1)}{v_1^2}	\\
		& = \frac{g(v_1)}{v_1}[2+\eta_g(v_1)]
\end{align*}
where $\eta_g$ is the elasticity of $g$. Hence, SC-H holds if and only if $\eta_g(v_1) \ge -2$ for all $v_1$.

For any $(v_1,v_2)$,\vspace{-3mm}
\begin{align*}
\Phi(v_1,v_2) & = f(1,v_2)-\int\limits_{v_1}^1\Big[3f(x,v_2)+(x,v_2)\cdot\nabla f(x,v_2)\Big] dx\\
 & =  g(1)-  \int\limits_{v_1}^1\Big[ 2\frac{g(x)}{x}+\frac{dg(x)}{dx}\Big]dx\\
& = g(v_1)-  2\int\limits_{v_1}^1\frac{g(x)}{x} dx
\end{align*}
As $\Phi(v_1,v_2)$ is independent of $v_2$, SC-V is satisfied for any density $g$.

For any $y \in [0,1]$,
\begin{align*}
\Phi(y,y) &= g(1)-  \int\limits_y^1\Big[3f(x,y) + (x,y) \cdot \nabla f(x,y) \Big] dx
\end{align*}
The bracketed term in the integral is independent of $y$ [as $f(x,y)=g(x)/x$] and is non-negative if SC-H holds. Hence, $\Phi(y,y)$ is  increasing in $y$ if SC-H holds.\hfill$\Box$
}

\bigskip\noindent
{\bf Proof of Theorem \ref{theo:det}:}
As $f$ satisfies SC-H and SC-V, by Proposition 2 there is an optimal mechanism which is semi-deterministic: \rfr{$(q,t) \equiv (\underline{t},q_2(1,a\alpha),\underline{v}_2,\bar{v}_2)$, where $\underline{t} \equiv t(1,0)$.} If $q_2(1,a\alpha)= 0$ or 1, then $(q,t)$ is deterministic. Therefore, assume that $q_2(1,a\alpha)\in(0,1)$. Figure \ref{fig:f7} shows such a semi-deterministic mechanism.\footnote{\rfr{One or more of the four regions in Figure \ref{fig:f7} may be empty.}}

\begin{figure}[!hbt]
  \centering
  \includegraphics[width=4.5in]{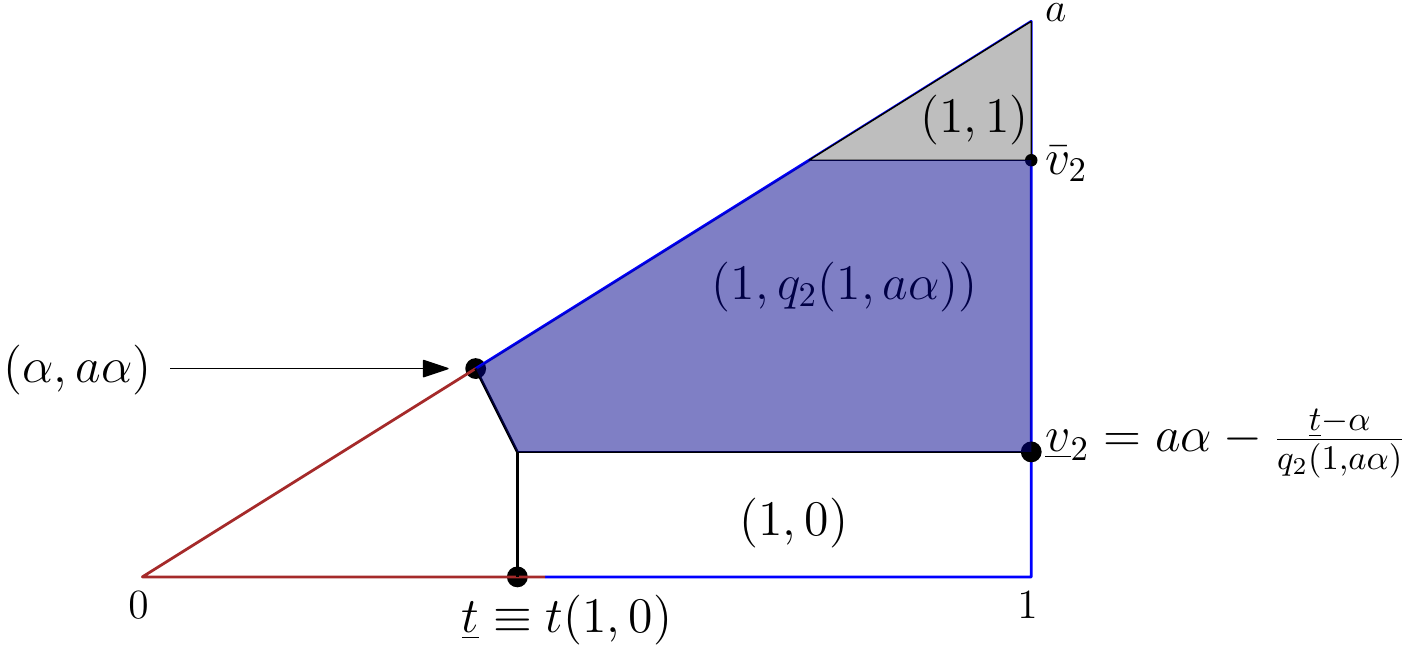}
  \caption{A semi-deterministic mechanism}
  \label{fig:f7}
\end{figure}

In this semi-deterministic mechanism,
\begin{eqnarray}
u(1,v_2) & = &
\begin{cases}
1- \underline{t}, & \textrm{if}~v_2 \le \underline{v}_2 \\ \label{eq:uti}
1 + (v_2 - \underline{v}_2)q_2(1,a\alpha) - \underline{t}, & \textrm{if}~v_2 \in [\underline{v}_2,a\alpha] \\
1- \alpha + \int \limits_{a\alpha}^{v_2}q_2(1,y)dy, & \textrm{if}~v_2 \in [a\alpha,a],
\end{cases}
\end{eqnarray}
where $q_2(1,y)=q_2(1,a\alpha)$ if $y \in [a\alpha,\bar{v}_2)$ and $q_2(1,y)=1$ if $y \in [\bar{v}_2,a]$.

From Lemma 4, we have
\begin{align}
  \textsc{Rev}(q,t) &= \int \limits_0^{a\alpha} \int \limits_{1-u(1,v_2)}^1 \Phi(v_1,v_2) dv_1 dv_2
  + \int \limits_{a\alpha}^a \int \limits_{\frac{v_2}{a}}^1 \Phi(v_1,v_2) dv_1 dv_2
  - \int \limits_{a\alpha}^a (1-\frac{v_2}{a}-u(1,v_2))\Phi(\frac{v_2}{a},v_2)dv_2 \nonumber \\
  &= \int \limits_0^{\underline{v}_2} \int \limits_{\underline{t}}^1 \Phi(v_1,v_2) dv_1 dv_2 + \int \limits_{\underline{v}_2}^{a\alpha} \int \limits_{\underline{t}-(v_2-\underline{v}_2)q_2(1,a\alpha)}^1 \Phi(v_1,v_2) dv_1 dv_2 + \int \limits_{a\alpha}^a \int \limits_{\frac{v_2}{a}}^1 \Phi(v_1,v_2) dv_1 dv_2 \nonumber \\
  & \quad - \int \limits_{a\alpha}^a \Big(1-\frac{v_2}{a}\Big)\Phi(\frac{v_2}{a},v_2)dv_2 + (1-\alpha)\int \limits_{a\alpha}^a \Phi(\frac{v_2}{a},v_2)dv_2 +
  \int \limits_{a\alpha}^a \Big[\int \limits_{a\alpha}^{v_2}q_2(1,y)dy\Big] \Phi(\frac{v_2}{a},v_2)dv_2 \nonumber \\
  & \textrm{\hglue 2in [Inserting  $u(1,v_2)=1-\alpha+ \int_{a\alpha}^{v_2}q_2(1,y)dy$ from (\ref{eq:uti})]} \nonumber \\[5pt]
  &= \int \limits_0^{\underline{v}_2} \int \limits_{\underline{t}}^1 \Phi(v_1,v_2) dv_1 dv_2 + \int \limits_{\underline{v}_2}^{a\alpha} \int \limits_{\underline{t}-(v_2-\underline{v}_2)q_2(1,a\alpha)}^1 \Phi(v_1,v_2) dv_1 dv_2 + \int \limits_{a\alpha}^a \int \limits_{\frac{v_2}{a}}^1 \Phi(v_1,v_2) dv_1 dv_2 \nonumber \\
  & \quad - \int \limits_{a\alpha}^a \Big(1-\frac{v_2}{a}\Big)\Phi(\frac{v_2}{a},v_2)dv_2 + (1-\alpha)\int \limits_{a\alpha}^a \Phi(\frac{v_2}{a},v_2)dv_2 +
  \int \limits_{a\alpha}^a \Big[ \int \limits_{v_2}^{a}\Phi(\frac{y}{a},y)dy\Big] q_2(1,v_2) dv_2 \nonumber \\
  & \textrm{\hglue 2in [Changing the order of integration in the last term]} \nonumber \\[5pt]
  &= \int \limits_0^{\underline{v}_2} \int \limits_{\underline{t}}^1 \Phi(v_1,v_2) dv_1 dv_2 + \int \limits_{\underline{v}_2}^{a\alpha} \int \limits_{\underline{t}-(v_2-\underline{v}_2)q_2(1,a\alpha)}^1 \Phi(v_1,v_2) dv_1 dv_2 + \int \limits_{a\alpha}^a \int \limits_{\frac{v_2}{a}}^1 \Phi(v_1,v_2) dv_1 dv_2
  \nonumber \\
  & \quad - \int \limits_{a\alpha}^a \Big(1-\frac{v_2}{a}\Big)\Phi(\frac{v_2}{a},v_2)dv_2 + (1-\alpha)\int \limits_{a\alpha}^a \Phi(\frac{v_2}{a},v_2)dv_2 +
 q_2(1,a\alpha)  \int \limits_{a\alpha}^{\bar{v}_2} \Big[ \int \limits_{v_2}^{a}\Phi(\frac{y}{a},y)dy\Big]dv_2 \nonumber \\
   &   \quad + \int \limits_{\bar{v}_2}^a \Big[ \int \limits_{v_2}^{a}\Phi(\frac{y}{a},y)dy\Big]dv_2 \label{eq:revsemi}
\end{align}

As noted at the beginning of the proof, $q_2(1,a\alpha)<1$. Then $\bar{v}_2 > a\alpha$. Differentiate $\textsc{Rev}(q,t)$ with respect to $q_2(1,a\alpha)$, changing $\alpha$ but not changing $\bar{v}_2,\underline{v}_2$, and $\underline{t}$. As $q_2(1,a\alpha)<~1$, we know from Definition~\ref{d:clm} that $q_2(1,a\alpha)=\bar{q}_2$. Note that $\frac{d\alpha}{d\bar{q}_2}=-\frac{a\alpha - \underline{v}_2}{1+a\bar{q}_2}$. Using this and (\ref{eq:revsemi}),
we have
\begin{align*}
\frac{\partial \textsc{Rev}(q,t)}{\partial \bar q_2} &= a\frac{d\alpha}{d\bar{q}_2} \int \limits_{\alpha}^1 \Phi(v_1,a\alpha)dv_1 + \int \limits_{\underline{v}_2}^{a\alpha}(v_2-\underline{v}_2) \Phi\Big(\underline{t} - (v_2 - \underline{v}_2q_2(1,a\alpha),v_2\Big)dv_2 \\
&  \quad - a\frac{d\alpha}{d\bar{q}_2} \int \limits_{\alpha}^1 \Phi(v_1,a\alpha)dv_1 + a\frac{d\alpha}{d\bar{q}_2} (1-\alpha)\Phi(\alpha,a\alpha) - a\frac{d\alpha}{d\bar{q}_2} (1-\alpha) \Phi(\alpha,a\alpha) \\
&  \quad - \frac{d\alpha}{d\bar{q}_2} \int \limits_{a\alpha}^a \Phi(\frac{v_2}{a},v_2)dv_2
- aq_2(1,a\alpha)\frac{d\alpha}{d\bar{q}_2}  \int \limits_{a\alpha}^a \Phi(\frac{v_2}{a},v_2)dv_2
+ \int \limits_{a\alpha}^{\bar{v}_2} \Big[ \int \limits_{v_2}^{a}\Phi(\frac{y}{a},y)dy\Big] dv_2 \\
&= \int \limits_{\underline{v}_2}^{a\alpha}(v_2-\underline{v}_2) \Phi\Big(\underline{t} - (v_2 - \underline{v}_2)q_2(1,a\alpha),v_2\Big)dv_2 + (a\alpha-\underline{v}_2) \int \limits_{a\alpha}^a \Phi(\frac{v_2}{a},v_2)dv_2 \\
&  \quad + \int \limits_{a\alpha}^{\bar{v}_2} \Big[ \int \limits_{v_2}^{a}\Phi(\frac{y}{a},y)dy\Big] dv_2
\end{align*}

The first-order condition $\frac{\partial \textsc{Rev}(q,t)}{\partial \bar{q}_2}=0$ implies
\begin{align}
\int \limits_{a\alpha}^{\bar{v}_2} \Big[ \int \limits_{v_2}^{a}\Phi(\frac{y}{a},y)dy\Big] dv_2 &= -\int \limits_{\underline{v}_2}^{a\alpha}(v_2-\underline{v}_2) \Phi\Big(\underline{t} - (v_2 - \underline{v}_2)q_2(1,a\alpha),v_2\Big)dv_2 - (a\alpha-\underline{v}_2) \int \limits_{a\alpha}^a \Phi(\frac{v_2}{a},v_2)dv_2 \label{eq:nfoc1}
\end{align}
Consider two cases.

\medskip

\noindent \underline{\sc Case 1:} Suppose $\int \limits_{a\alpha}^a \Phi(\frac{v_2}{a},v_2)dv_2 \ge 0$. Then, SC-D implies that
$\int \limits_{v_2}^{a}\Phi(\frac{y}{a},y)dy \ge 0$ for all $v_2 > a\alpha$. Hence, we get
\begin{align}
\int \limits_{a\alpha}^{\bar{v}_2} \Big[ \int \limits_{v_2}^{a}\Phi(\frac{y}{a},y)dy\Big] dv_2 &\ge 0 \label{eq:nfoc2}
\end{align}

\noindent \underline{\sc Case 2:} Suppose $\int \limits_{a\alpha}^a \Phi(\frac{v_2}{a},v_2)dv_2 < 0$.
Lemma~\ref{lem:bound} implies that $\Phi(\underline{t},\underline{v}_2) \le 0$. Hence, SC-H and SC-V imply that $\Phi\Big(\underline{t} - (v_2 - \underline{v}_2)q_2(1,a\alpha),v_2\Big) \le 0$ for all
$v_2 \in [\underline{v}_2,a\alpha]$. This implies that the right-hand side of (\ref{eq:nfoc1}) is positive.
Hence, we have
\begin{align*}
\int \limits_{a\alpha}^{\bar{v}_2} \Big[ \int \limits_{v_2}^{a}\Phi(\frac{y}{a},y)dy\Big] dv_2 &> 0 
\end{align*}

So, in both cases, (\ref{eq:nfoc2}) holds.
Define a new mechanism $(q',t')$ from $(q,t)$ by increasing $q_2'(1,v_2)$ from $q_2(1,a\alpha)<1$ to 1 for all $v_2\in[a\alpha,\bar{v}_2)$. Thus, $\bar{v}_2'=a\alpha$ and everything else, including $\alpha, \underline{v}_2,$ and $\underline{t}$, remains as in $(q,t)$.  This only changes the last two terms of
(\ref{eq:revsemi}) and hence, we have
\begin{align*}
   \textsc{Rev}(q',t') -  \textsc{Rev}(q,t) & =    \int \limits_{a\alpha}^{\bar{v}_2} \Big[ \int \limits_{v_2}^{a}\Phi(\frac{y}{a},y)dy\Big] dv_2  - q_2(1,a\alpha) \int \limits_{a\alpha}^{\bar{v}_2} \Big[ \int \limits_{v_2}^{a}\Phi(\frac{y}{a},y)dy\Big]  dv_2\ \ge\ 0,
 \end{align*}
 where the inequality follows from (\ref{eq:nfoc2}). Thus, the revenue from $(q',t')$ is no less than that from $(q,t)$, \rfr{which was assumed to be optimal.
Hence, $(q',t')$ is an optimal mechanism. Note that it is semi-deterministic and $t'(1,0)=\underline{t}$.

  Let $(q^*,t^*)$ be the cover of $(q',t')$. By Lemma~\ref{le:covv}, $(q^*,t^*)$ is semi-deterministic and $q^*_2(1,a\alpha)=q_2'(1,a\alpha)$, which is 1 by assumption. Therefore, Definition~\ref{def:sdm} implies that $(q^*,t^*)$ is a deterministic mechanism. As $(q',t')$ is an optimal semi-deterministic mechanism, Lemma~\ref{lem:bound} implies that $\Phi(\underline{t},\underline{v}_2) \le 0$. Hence, by Lemma~\ref{lem:cov1},  $(q^*,t^*)$ generates at least as much revenue as $(q',t')$. Hence, $(q^*,t^*)$ is optimal, and as noted earlier, it is deterministic.}
\hfill$\blacksquare$

\subsection{Proof of Section~\ref{sec:nc}}\label{prf:nc}

\noindent
{\bf Proof of Proposition~\ref{prop:nec}:}
A deterministic mechanism is a constrained line mechanism.  Therefore, use (\ref{eq:rline1}) and~(\ref{eq:rline2}) to obtain the expected revenue from prices $(p_1,p_2)$:
\begin{eqnarray} \nonumber
\textsc{Rev}(p_1,p_2) &= &\int \limits_0^{p_2} \int \limits_{p_1}^1 \Phi(v_1,v_2)dv_1 dv_2 +
\int \limits_{p_2}^{a\alpha} \int \limits_{(1+a)\alpha - v_2}^1 \Phi(v_1,v_2)dv_1 dv_2 \\ \nonumber
&& \quad + \ \int \limits_{a\alpha}^a\int \limits_{\frac{v_2}a}^1 \Phi(v_1,v_2) dv_1 dv_2 +\frac{1+a}a\int \limits_{a\alpha}^a (v_2-a\alpha) \Phi(\frac{v_2}a,v_2)dv_2
\end{eqnarray}
where we use  $u(1,v_2)=1+v_2-(1+a)\alpha$ for $v_2\ge a\alpha$.

As $(1+a)\alpha=p_1+p_2$, we have
\begin{align*}
\frac{d\alpha}{dp_1} &=\frac{d\alpha}{dp_2} = \frac{1}{1+a}
\end{align*}
The derivatives are well-defined as $p_1$ and $p_2$ are in the interior of the domain.
The first-order conditions at optimal prices $(p_1^*,p_2^*)$ are
\begin{eqnarray*}
\frac{\partial\textsc{Rev}(q,t)}{\partial p_1}  & = &-\int \limits_0^{p_2^*} \Phi(p_1,v_2) dv_2 - \int \limits_{p_2^*}^{a\alpha^*}\Phi((1+a)\alpha-v_2,v_2)dv_2 -\int_{a\alpha^*}^a \Phi(\frac{v_2}a,v_2)dv_2  \ =\ 0 \\[5pt]
\frac{\partial \textsc{Rev}(q,t)}{\partial p_2}
& = & - \int \limits_{p_2^*}^{a\alpha^*}\Phi((1+a)\alpha^*-v_2,v_2)dv_2 -\int_{a\alpha^*}^a\Phi(\frac{v_2}a,v_2)dv_2 \ = \ 0
\end{eqnarray*}
The second equation is (\ref{eq:nece2}). Inserting it in the first equation above yields (\ref{eq:nece1}).

That  $\Phi(p_1^*,p_2^*)\le 0$ follows from Lemma 5 as $p_1^*=t(1,0)$ and $p_2^*=\underline{v}_2$.

Finally, suppose that $\Phi(p_1^*,0)< 0$. Then by SC-V, $\Phi(p_1^*,v_2)\le0$ for all $v_2$. By continuity, there is an $\epsilon > 0$ such that $\Phi(p_1^*,v_2)< 0$ for all $v_2 <\epsilon$. Consequently, (\ref{eq:nece1}) is not satisfied. Hence, $\Phi(p_1^*,0)\ge 0$. \hfill$\blacksquare$

\subsection{Proofs of Sections~\ref{sec:odvm} and \ref{sec:sdvm}}\label{prf:odvm}

\noindent
{\bf Proof of Proposition \ref{prop:orddec}:}
 {\sc SC-H:} For any $(v_1,v_2)$,
\begin{align}\nonumber
a[3f(v) + v \cdot \nabla f(v)] &= 6 g(v_1)g(\frac{v_2}{a}) + 2v_1 \frac{dg(v_1)}{dv_1} g(\frac{v_2}{a}) + 2v_2 \frac{dg(\frac{v_2}{a})}{dv_2} g(v_1) \\ \nonumber
&= 2g(v_1)g(\frac{v_2}{a}) \Big[3 + \frac{v_1}{g(v_1)} \frac{dg(v_1)}{dv_1} + \frac{v_2}{g(\frac{v_2}{a})} \frac{dg(\frac{v_2}{a})}{dv_2}\Big] \\ \label{eq:sch}
&= 2g(v_1)g(\frac{v_2}{a}) \Big[3+\eta_g(v_1)+\eta_g(\frac{v_2}{a})\Big]
\end{align}
Hence, if $\eta_g(x) \ge -\frac{3}{2}$ for all $x$, then SC-H holds. If, instead, $\eta_g(x)<-\frac32$ for some $x$, then SC-H is violated at $v_1=\frac{v_2}a = x$.\\

\noindent {\sc SC-V:}
For any $(v_1,v_2)$,\vspace{-3mm}
\begin{eqnarray}\nonumber
a\Phi(v_1,v_2) & = & 2g(1)g(\frac{v_2}{a})- 2g(\frac{v_2}{a})\int\limits_{v_1}^1\Big[3g(x)+x \frac{dg(x)}{dx}+\eta_g(\frac{v_2}{a}) g(x)\Big]dx\\\nonumber
& = & 2v_1g(v_1)g(\frac{v_2}{a}) -2g(\frac{v_2}{a})\Big(1-G(v_1)\Big)\Big(2+\eta_g(\frac{v_2}{a})\Big)\\\nonumber
&= & 2 g(v_1) g(\frac{v_2}{a}) \Bigg[v_1 - \frac{1-G(v_1)}{g(v_1)} \Big(2+\eta_g(\frac{v_2}{a})\Big)\Bigg]\\\label{eq:vu}
&= & 2 g(v_1) g(\frac{v_2}{a}) W(v_1,v_2) \label{eq:phi1}
\end{eqnarray}
Thus, SC-V is equivalent to $W(v_1,\cdot)$ crosses zero at most once for all $v_1$. \\

\noindent {\sc SC-D:} For any $y \in [0,a]$,
\begin{align*}
a\Phi(\frac{y}{a},y) &= 2 g(\frac{y}{a}) g(\frac{y}{a}) \Bigg[\frac{y}{a} - \frac{1-G(\frac{y}{a})}{g(\frac{y}{a})} \Big(2+\eta_g(\frac{y}{a})\Big)\Bigg],
\end{align*}
Denoting $y':=\frac{y}{a}$, this simplifies to
\begin{align*}
a\Phi(\frac{y}{a},y) &= 2 [g(y')]^2 \Bigg[y' - \frac{1-G(y')}{g(y')} \Big(2+\frac{1}{g(y')}y' \frac{dg(y')}{dy'}\Big)\Bigg] \\
&= 2y' [g(y')]^2 - 4(1-G(y'))g(y') - 2y'(1-G(y'))\frac{dg(y')}{dy'} \\
&= -2 \frac{d[y'g(y')(1-G(y'))]}{dy'} + \frac{d(1-G(y'))^2}{dy'},  \qquad \forall y'\in [0,1]
\end{align*}

Integrating both sides,
\begin{align*}
a \int \limits_{\frac{v_2}{a}}^1 \Phi(y',ay') dy' &= 2\frac{v_2}{a}g(\frac{v_2}{a})(1-G(\frac{v_2}{a})) - (1-G(\frac{v_2}{a}))^2 \\
&= v_2 g_{min}(v_2) - (1-G_{min}(v_2)) \\
&= g_{min}(v_2) \Big[v_2 - \frac{1-G_{min}(v_2)}{g_{min}(v_2)}\Big] \\
&= a^2 g_{min}(v_2)W_{min}(v_2)
\end{align*}
Hence,
\begin{align}\label{eq:phi2}
 \int \limits_{v_2}^a \Phi(\frac{y}{a},y) dy &= a g_{min}(v_2)W_{min}(v_2).
\end{align}
Thus, SC-D is equivalent to the requirement that $W_{min}$ crosses zero at most once.

By Theorem \ref{theo:det}, the existence of an optimal mechanism that is deterministic follows.
\hfill$\blacksquare$

\medskip

\noindent {\bf Proof of Proposition~\ref{prop:ordvu}:} As $(p_1,p_2)$ is deterministic, we have
\begin{align*}
u(1,v_2) =
\begin{cases}
1-p_1 & \textrm{if}~v_2 \le p_2 \\
1+v_2 - p_1 - p_2 & \textrm{if}~v_2 > p_2.
\end{cases}
\end{align*}
Since a deterministic mechanism is a line mechanism, Lemma~\ref{lem:revline} implies that
\begin{align*}
\textsc{Rev}(p_1,p_2) &= \int \limits_0^{a\alpha} \int \limits_{1-u(1,v_2)}^1 \Phi(v_1,v_2)dv_1dv_2 + \int \limits_{a\alpha}^a \int \limits_{\frac{v_2}{a}}^1 \Phi(v_1,v_2)dv_1dv_2 \\
&\qquad +\ \int \limits_{a\alpha}^a \big(u(1,v_2) - (1 - \frac{v_2}{a})\big)\Phi(\frac{v_2}{a},v_2)dv_2
\end{align*}
We use the following in the proof:
\begin{align}\label{eq:prf}
\int\limits_{x}^a \bigg[\int\limits_{v_2}^a\Phi(\frac y a ,y)dy\bigg]dv_2 =
\int\limits_{x}^a \bigg[\int\limits_{x}^{y}dv_2\bigg]\Phi(\frac{y}{a},y)dy = \int\limits_{x}^a (y-x) \Phi(\frac{y}a,y)dy =\int\limits_{x}^a (v_2-x) \Phi(\frac{v_2}a, v_2)dv_2
\end{align}
Consider two cases: \\

\noindent \underline{\sc Case 1: $ap_1 \ge p_2$.} In this case, $p_1+p_2=(1+a)\alpha$ and $S_1^d(p_1,p_2)$ has zero measure. Hence,
\begin{align*}
\textsc{Rev}(p_1,p_2) &= \int \limits_0^{p_2} \int \limits_{p_1}^1 \Phi(v_1,v_2)dv_1dv_2 +
\int \limits_{p_2}^{a\alpha} \int \limits_{p_1+p_2 - v_2}^1 \Phi(v_1,v_2)dv_1dv_2 \\
&\qquad + \int \limits_{a\alpha}^a \int \limits_{\frac{v_2}{a}}^1 \Phi(v_1,v_2)dv_1dv_2 + \int \limits_{a\alpha}^a \big(\frac{1+a}{a}v_2- (p_1+p_2)\big)\Phi(\frac{v_2}{a},v_2)dv_2 \\
&= \int \limits_0^{p_2} \int \limits_{p_1}^1 \Phi(v_1,v_2)dv_1dv_2 +
\int \limits_{p_2}^{a\alpha} \int \limits_{(1+a)\alpha - v_2}^1 \Phi(v_1,v_2)dv_1dv_2 \\
&\qquad + \int \limits_{a\alpha}^a \int \limits_{\frac{v_2}{a}}^1 \Phi(v_1,v_2)dv_1dv_2 + \frac{1+a}{a} \int \limits_{a\alpha}^a (v_2-a\alpha)\Phi(\frac{v_2}{a},v_2)dv_2,
\end{align*}
where we used $p_1+p_2=(1+a)\alpha$ in the last step.
Then $\Phi(v_1,v_2)=W(v_1,v_2)f(v_1,v_2)$ (from (\ref{eq:phi1})) and $\int \limits_{v_2}^a \Phi(\frac{y}{a},y)dy =aW_{min}(v_2) g_{min}(v_2) $ (from (\ref{eq:phi2})), together with (\ref{eq:prf}), imply that
\begin{align*}
\textsc{Rev}(p_1,p_2) &=\int \limits_0^{p_2} \int \limits_{p_1}^1 W(v_1,v_2)f(v_1,v_2)dv_1 dv_2 +
\int \limits_{p_2}^{a\alpha} \int \limits_{(1+a)\alpha - v_2}^1W(v_1,v_2)f(v_1,v_2)dv_1 dv_2 \\
& \qquad + \ \int \limits_{a\alpha}^a\int \limits_{\frac{v_2}a}^1W(v_1,v_2)f(v_1,v_2) dv_1 dv_2 +(1+a)\int\limits_{a\alpha}^a W_{min}(v_2) g_{min}(v_2) dv_2\\
& = \e\Big[W(v_1,v_2)1_{\big\{ S_1(p_1,p_2) \big\}}\Big] +(1+a)\e\Big[W_{min}(v_2)1_{\big\{ S^d_2(p_1,p_2)\big\}}\Big]
\end{align*}
Since $S_1^d(p_1,p_2)=\emptyset$, the above expression is equivalent to (\ref{eq:ordvu}). \\

\noindent \underline{\sc Case 2: $ap_1 < p_2$.} In this case, $\alpha=p_1$ and $S_1^d(p_1,p_2)$ is nonempty.
Hence,
\begin{align*}
\textsc{Rev}(p_1,p_2) &= \int \limits_0^{ap_1} \int \limits_{p_1}^1 \Phi(v_1,v_2)dv_1dv_2 + \int \limits_{ap_1}^a \int \limits_{\frac{v_2}{a}}^1 \Phi(v_1,v_2)dv_1dv_2 + \int \limits_{ap_1}^{p_2}\big(\frac{v_2}{a} - p_1 \big)\Phi(\frac{v_2}{a},v_2)dv_2 \\
&\qquad + \int \limits_{p_2}^{a}\big(\frac{v_2}{a}- p_1+ v_2  - p_2) \big)\Phi(\frac{v_2}{a},v_2)dv_2 \\
&= \int \limits_0^{ap_1} \int \limits_{p_1}^1 \Phi(v_1,v_2)dv_1dv_2 + \int \limits_{ap_1}^a \int \limits_{\frac{v_2}{a}}^1 \Phi(v_1,v_2)dv_1dv_2 + \int \limits_{ap_1}^{a}\big(\frac{v_2}{a} - p_1\big)\Phi(\frac{v_2}{a},v_2)dv_2 \\
&\qquad + \int \limits_{p_2}^{a}\big(v_2 - p_2 \big)\Phi(\frac{v_2}{a},v_2)dv_2
\end{align*}

Using (\ref{eq:phi1}), (\ref{eq:phi2}) and (\ref{eq:prf}) we have
\begin{align*}
\textsc{Rev}(p_1,p_2) &= \int \limits_0^{ap_1} \int \limits_{p_1}^1 \Phi(v_1,v_2)dv_1dv_2 + \int \limits_{ap_1}^a \int \limits_{\frac{v_2}{a}}^1 \Phi(v_1,v_2)dv_1dv_2 + \frac{1}{a}\int \limits_{ap_1}^a \Big(\int \limits_{v_2}^a \Phi(\frac{y}{a},y)dy \Big)dv_2 \\
&\qquad + \int \limits_{p_2}^a \Big(\int \limits_{v_2}^a \Phi(\frac{y}{a},y)dy \Big)dv_2 \\
&= \int \limits_0^{ap_1} \int \limits_{p_1}^1 \Phi(v_1,v_2)dv_1dv_2 + \int \limits_{ap_1}^a \int \limits_{\frac{v_2}{a}}^1 \Phi(v_1,v_2)dv_1dv_2 + \frac{1}{a}\int \limits_{ap_1}^{p_2} \Big(\int \limits_{v_2}^a \Phi(\frac{y}{a},y)dy \Big)dv_2 \\
&\qquad + \frac{1+a}{a} \int \limits_{p_2}^a \Big(\int \limits_{v_2}^a \Phi(\frac{y}{a},y)dy \Big)dv_2 \\
&=\int \limits_0^{ap_1} \int \limits_{p_1}^1 W(v_1,v_2)f(v_1,v_2) dv_1dv_2 + \int \limits_{ap_1}^a \int \limits_{\frac{v_2}{a}}^1 W(v_1,v_2) f(v_1,v_2) dv_1dv_2 \\
&\qquad + \int \limits_{ap_1}^{p_2} W_{min}(v_2) g_{min}(v_2) dv_2
+ (1+a) \int \limits_{p_2}^a W_{min}(v_2) g_{min}(v_2)dv_2 \\
&= \e\Big[W(v_1,v_2)1_{\big\{ S_1(p_1,p_2)\big\}}\Big] + \e\Big[W_{min}(v_2) 1_{\big\{S^d_1(p_1,p_2) \big \}}\Big] + (1+a)\e\Big[W_{min}(v_2) 1_{\big\{S^d_2(p_1,p_2) \big \}}\Big]
\end{align*}

\hfill$\blacksquare$

\vspace{0.1in}

\noindent
\aed{
{\bf Proof of Proposition \ref{prop:sdvm2}:}\\
SC-H: Note that
\begin{align*}
\nabla f(v_1,v_2) & = \Bigg(\frac{dg_1(v_1)}{dv_1}\frac{g_2(v_2)}{1-G_1(v_2)},\ g_1(v_1)\frac{\frac{d[g_2(v_2)]}{dv_2}(1-G_1(v_2))+g_1(v_2)g_2(v_2)}{(1-G_1(v_2))^2} \Bigg)\\[5pt]
3f(v)+v\cdot \nabla f(v) & = 3\frac{g_1(v_1)}{1-G_1(v_2)}g_2(v_2)+\frac{g_1(v_1)}{1-G_1(v_2)}g_2(v_2)\Big[ \eta_{g_1}(v_1) +\eta_{g_2}(v_2)+v_2\frac{g_1(v_2)}{1-G_1(v_2)}\Big] \\
& = f(v_1.v_2)\Bigg(3+ \eta_{g_1}(v_1) +\eta_{g_2}(v_2)+v_2\frac{g_1(v_2)}{1-G_1(v_2)} \Bigg)
\end{align*}
As $v_2\frac{g_1(v_2)}{1-G_1(v_2)}\ge 0$ for all $v_2$, SC-H is satisfied if $\eta_{g_1}(v_1)+\eta_{g_2}(v_2)\ge -3$, $\forall v_1\ge v_2$.\\

\noindent
{\sc SC-V:}
To simplify the notation, for every $v_2$ let
\begin{align*}
\tau(v_2) &:= \frac{g_2(v_2)}{1-G_1(v_2)} \\
\gamma(v_2) &:= \eta_{g_2}(v_2) +v_2\frac{g_1(v_2)}{1-G_1(v_2)}
\end{align*}
Note $f(v_1,v_2)=g_1(v_1)\tau(v_2)$. Then,
\begin{align*}
\int \limits_{v_1}^1 \Big[ 3f(x,v_2) + (x,v_2) \cdot \nabla f(x,v_2)\Big]dx &=
\tau(v_2) \int \limits_{v_1}^1 \Big[ 3g_1(x) + x\frac{d[g_1(x)]}{dx} + g_1(x)\gamma(v_2)\Big] dx \\
&= \tau(v_2) \int \limits_{v_1}^1 \Big[ \frac{d[xg_1(x)]}{dx} + g_1(x)\big(2+\gamma(v_2)\big)\Big] dx \\
&= \tau(v_2) \Big[ g_1(1) - v_1 g_1(v_1) + (1-G_1(v_1))(2+\gamma(v_2))\Big]
\end{align*}
Thus,\vspace{-3mm}
\begin{align*}
\Phi(v_1,v_2) & = f(1,v_2)-\int\limits_{v_1}^1\Big[3f(x,v_2)+(x,v_2)\cdot\nabla f(x,v_2)\Big] dx\\
& = g_1(1)\tau(v_2)- \tau(v_2) \Big[ g_1(1) - v_1 g_1(v_1) + (1-G_1(v_1))(2+\gamma(v_2))\Big]\\
 & = \tau(v_2) \Big[ v_1g_1(v_1) - (1-G_1(v_1))(2+\gamma(v_2))\Big]
 \end{align*}
Note that $\tau$ is positive. Hence, $\Phi$ satisfies SC-V if $\gamma$ is increasing and
$2+\gamma(v_2) \ge 0$. The condition $\eta_{g_2}(v_2) \ge -2$ implies that $2+\gamma(v_2) \ge 0$.

\noindent {\sc SC-D:} For any $y \in [0,1]$,
\begin{align*}
\Phi(y,y) &= \tau(y) \Big[ yg_1(y) - (1-G_1(y))(2+\gamma(y))\Big] \\
&= g_2(y) \Big[ y\frac{g_1(y)}{1-G_1(y)} - 2 - \eta_{g_2}(y) - y \frac{g_1(y)}{1-G_1(y)}\Big] \\
&= -g_2(y) (2+\eta_{g_2}(y))
\end{align*}
So, $\eta_{g_2} \ge -2$ ensures SC-D.

By Theorem \ref{theo:det}, if (i), (ii), and (iii) are satisfied, then there is an optimal mechanism that is deterministic.
\hfill$\blacksquare$
}

\subsection{Proofs of Section \ref{sec:imv}}\label{prf:imv}

\noindent
{\bf Proof of Proposition~\ref{prop:2}:}
Given an IC and IR mechanism $(q,t)$, construct another mechanism $(\hat{q},\hat{t})$ as follows.
Let $X:=\{(v_1,1): v_1 \in [0,a]\}$. We first define $(\hat{q},\hat{t})$ on $X$.
For each $v_1 \in [0,a]$, let
\begin{eqnarray*}
\hat{q}_1(v_1,1) &:= &  q_1(v_1,1),\qquad \hat{q}_2(v_1,1) \ := \ q_1(v_1,1) \\
\hat{t}(v_1,1) &:= &t(v_1,1) + \big[q_1(v_1,1)-q_2(v_1,1)\big]
\end{eqnarray*}
Thus, we keep the allocation probability of the first unit unchanged and increase the allocation probability of the second unit to the maximum feasible. For every $v \in D \setminus X$, set $\hat{q}_2(v):=\hat{q}_1(v)$ and
\begin{align*}
\Big(\hat{q}_1(v),\hat{t}(v)\Big) :=
\begin{cases}
(0,0), & \textrm{if}~(v_1+v_2)\hat{q}_1(0,1) < t(0,1) \\
\big(\hat{q}_1(0,1),\hat{t}(0,1)\big), & \textrm{if}~v_1+v_2 < 1, (v_1+v_2)\hat{q}_1(0,1) \ge \hat{t}(0,1) \\
\big(\hat{q}_1(v_1+v_2-1,1),\hat{t}(v_1+v_2-1)\big), & \textrm{otherwise}
\end{cases}
\end{align*}

We first show that $(\hat{q},\hat{t})$ restricted to $X$ is IC and IR.
Note that for all $(v_1,1) \in X$,
\begin{align}\label{eq:imv1}
\hat{u}(v_1,1) &= (v_1+1)q_1(v_1,1) - t(v_1,1) - \big[q_1(v_1,1)-q_2(v_1,1)\big] = u(v_1,1)
\end{align}
Hence, IR of $(q,t)$ implies IR of $(\hat{q},\hat{t})$ restricted to $X$.
Similarly, IC of $(q,t)$ implies for every $(v_1,1),(v'_1,1) \in X$, we have
\begin{align*}
\hat{u}(v_1,1) - \hat{u}(v'_1,1) &= u(v_1,1) - u(v'_1,1) \ge q_1(v'_1,1) (v_1-v'_1) = \hat{q}_1(v'_1,1)(v_1-v'_1)
\end{align*}

Next, we show that $(\hat{q},\hat{t})$ is IC and IR of on $D \setminus X$. Note that the range of this mechanism
is $(0,0,0)$ and outcomes on $X$.  The payoff of type $(v_1,v_2) \in D \setminus X$ from the outcome for type $(v'_1,1) \in X$ is
\begin{align}
(v_1+v_2)\hat{q}_1(v'_1,1) &- \hat{t}(v'_1,1) = (v_1+v_2-v'_1-1)\hat{q}_1(v'_1,1) + \hat{u}(v'_1,1) \label{eq:imv2}\\
&= (v_1+v_2-1)\hat{q}_1(v'_1,1) - v'_1 \hat{q}_1(v'_1,1) + \hat{u}(0,1) + \int \limits_0^{v'_1}\hat{q}_1(x,1)dx \nonumber \\
&= (v_1+v_2-1)\hat{q}_1(v'_1,1) + \hat{u}(0,1) - \int \limits_0^{v'_1}\big[\hat{q}_1(v'_1,1)-\hat{q}_1(x,1)\big]dx \label{eq:imv3}
\end{align}
where the second equality follows from IC of $(\hat{q},\hat{t})$ on $X$. Now, consider two cases for $(v_1,v_2) \in D \setminus X$. \medskip

\noindent {\sc Case 1:} If $v_1+v_2 - 1 < 0$, then eq. (\ref{eq:imv3}) is maximized
at $v'_1=0$ (as IC of $(q,t)$ implies that $\hat q_1(v_1',1)=q_1(v_1',1)$ is increasing in $v_1'$). The payoff of type $(v_1,v_2)$ from the outcome for type $(0,1)$ is
$$(v_1+v_2-1)\hat{q}_1(0,1) + \hat{u}(0,1) = (v_1+v_2)\hat{q}_1(0,1) - \hat{t}(0,1)$$
If $(v_1+v_2)\hat{q}_1(0,1) < \hat{t}(0,1)$, type $(v_1,v_2)$ prefers the outcome $(0,0,0)$ to the outcome for type $(0,1)$.
If $v_1+v_2 < 1$ and $(v_1+v_2)\hat{q}_1(0,1) \ge \hat{t}(0,1)$, then type $(v_1,v_2)$ prefers outcome for $(0,1)$ to
every other outcome on $X$ and $(0,0,0)$. Thus, if $v_1+v_2 < 1$, type $(v_1,v_2) \in D \setminus X$ cannot manipulate $(\hat{q},\hat{t})$ and IR also holds. \medskip

\noindent {\sc Case 2:} Consider the case when $v_1+v_2 \ge 1$. With $\bar{v}_1 \equiv v_1+v_2-1$, we have
\begin{align*}
\hat{u}(\bar{v}_1,1) &= u(\bar{v}_1,1) \ge u(v'_1,1) + (\bar{v}_1-v'_1)\hat{q}_1(v'_1,1)~\qquad~\forall~v'_1 \in [0,a],
\end{align*}
where the equality follows from (\ref{eq:imv1}) and the inequality follows from IC of $(q,t)$.
From (\ref{eq:imv2}), $u(v'_1,1) + (\bar{v}_1-v'_1)\hat{q}_1(v'_1,1)$ is the payoff of type $(v_1,v_2)$
from the outcome for type $(v'_1,1)$. Hence, the outcome for $(\bar{v}_1,1)$ maximizes the payoff of
$(v_1,v_2)$ among all outcomes of types in $X$. Thus, $(v_1,v_2)$ cannot manipulate $(\hat{q},\hat{t})$. \medskip

So, $(\hat{q},\hat{t})$ is IC and IR.

Finally, for every $(v_1,v_2) \in D \setminus X$ if $\hat{u}(v_1,v_2) = 0$ then $u(v_1,v_2) \ge \hat{u}(v_1,v_2)$
due to IR of $(q,t)$. Else, the outcome of $(v_1,v_2)$ is the same as the outcome of type $(v'_1,1) \in X$, where
$v'_1 = \max(v_1+v_2-1,0)$.
IC of $(q,t)$ implies
\begin{align*}
u(v_1,v_2) &\ge u(v'_1,1) + (v_1-v'_1)q_1(v'_1,1) + (v_2-1)q_2(v'_1,1)  \\
&\ge u(v'_1,1) + (v_1-v'_1)q_1(v'_1,1) + (v_2-1)q_1(v'_1,1)  \\
&= (v_1+v_2-v'_1-1)\hat{q}_1(v'_1,1) + \hat{u}(v'_1,1) \\
&= \hat{u}(v_1,v_2),
\end{align*}
where the second inequality follows from $q_1(v'_1,1) \ge q_2(v'_1,1)$ and $v_2 < 1$, the first equality from (\ref{eq:imv1}), and the second equality from (\ref{eq:imv2}).

Hence, we have proved
\begin{eqnarray}\nonumber
\hat{u}(v_1,v_2) &= & u(v_1,v_2)~\qquad~\forall~(v_1,v_2) \in X \\ \label{eq:utless}
\hat{u}(v_1,v_2) &\le & u(v_1,v_2)~\qquad~\forall~(v_1,v_2) \in D \setminus X
\end{eqnarray}

Using integration by parts, as in the proof of Lemma~\ref{le:1}, one can show that:
\begin{align}
\textsc{Rev}(q,t) &= \int_0^a u(v_1,1)f(v_1,1)dv_1 - \int_0^1  \int_0^{av_2} u(v) \Big[ 3 f(v) + v \cdot \nabla f(v)\Big] dv_1dv_2 \label{eq:rev}
\end{align}
Since $f$ satisfies SC-H,  (\ref{eq:utless}) and (\ref{eq:rev}) imply that $\mbox{\sc Rev}(\hat q,\hat t) \ge \mbox{\sc Rev}(q,t)$.\hfill$\blacksquare$

\bigskip
\noindent
{\bf Proof of Proposition~\ref{pr:6}:}
With $v_1\le av_2$, we have
$$f(v_1,v_2)\ =\ \frac2a g(\frac{v_1}{a})g(v_2)$$
Analogous to the derivation of (\ref{eq:sch}), we have
\begin{align*}
a[3f(v_1,v_2)+(v_1,v_2)\cdot\nabla f(v_1,v_2)] & =  6g(\frac{v_1}{a})g(v_2)\Big[3+\eta_g(\frac{v_1}{a}) +\eta_g(v_2)\Big]
\end{align*}
Hence, if $\eta_g(x) \ge -\frac{3}{2}$ for all $x$, then SC-H is satisfied.

Theorem~4.5.8 in \cite{BP75} implies that $v_1, v_2$ have increasing hazard rates. That $v_1+v_2=X_1+X_2$ has increasing hazard rate follows from Corollary 1.B.39 in \cite{SS07}. Consequently, $w=v_1+v_2$ is regular.
\hfill$\blacksquare$


\newpage


\begin{thebibliography}{27}
\newcommand{\enquote}[1]{``#1''}
\expandafter\ifx\csname natexlab\endcsname\relax\def\natexlab#1{#1}\fi

\bibitem[\protect\citeauthoryear{Armstrong}{Armstrong}{1996}]{Ar96}
\textsc{Armstrong, M.} (1996): \enquote{Multiproduct Nonlinear Pricing,}
  \emph{Econometrica}, 64, 51--75.

\bibitem[\protect\citeauthoryear{Armstrong}{Armstrong}{2016}]{Ar16}
---\hspace{-.1pt}---\hspace{-.1pt}--- (2016): \enquote{Nonlinear Pricing,}
  \emph{Annual Review of Economics}, 8, 583--614.

\bibitem[\protect\citeauthoryear{Babaioff, Immorlica, Lucier, and Weinberg}{2020}]{BILW20}
  \textsc{Babaioff, M., N.~ Immorlica, B.~Lucier, and M.Weinberg} (2020):
   \enquote{A Simple and Approximately Optimal Mechanism for an Additive Buyer,}
  \emph{Journal of the ACM (JACM)}, 67, 4, 1--40.

\bibitem[\protect\citeauthoryear{Babaioff, Nisan, and Rubinstein}{2018}]{BNR18}
 \textsc{Babaioff, M., N.~Nisan, and A.~Rubinstein} (2018):
 \enquote{Optimal deterministic mechanisms for an additive buyer,}
 \emph{Proceedings of the 2018 ACM conference on Economics and Computation.}

\bibitem[\protect\citeauthoryear{Barlow and Proschan}{Barlow and
  Proschan}{1975}]{BP75}
\textsc{Barlow, R.~E. and F.~Proschan} (1975):
\emph{Statistical Theory of Reliability and Life Testing}, Holt, Rinehart, New York.

\bibitem[\protect\citeauthoryear{Bergemann, Brooks, and Morris}{Bergemann
  et~al.}{2020}]{BBM20}
\textsc{Bergemann, D., B.~Brooks, and S.~Morris} (2020):
\enquote{Countering the Winner's Curse: Optimal Auction Design in a Common Value Model,}
  \emph{forthcoming, Theoretical Economics}.

\bibitem[\protect\citeauthoryear{Bhattacharya,  Koutsoupias,  Kulkarni,  Leonardi,  Roughgarden,  and Xu}{Bhattacharya et al.}{2020}]{BKKLRX20}
 \textsc{Bhattacharya, S., E.~Koutsoupias, J.~Kulkarni, S.~Leonardi, T.~Roughgarden, X.,~Xu} (2020):
 \enquote{Prior-free Multi-unit Auctions with Ordered bidders.}
  \emph{forthcoming, Theoretical Computer Science}.

\bibitem[\protect\citeauthoryear{B{\"o}rgers}{B{\"o}rgers}{2015}]{Bo15}
\textsc{B{\"o}rgers, T.} (2015): \emph{An Introduction to the Theory of
  Mechanism Design}, Oxford University Press, USA.

\bibitem[\protect\citeauthoryear{Bulow and Klemperer}{Bulow and
  Klemperer}{2002}]{BK02}
\textsc{Bulow, J. and P.~Klemperer} (2002): \enquote{Prices and the Winner's
  Curse,} \emph{RAND Journal of Economics}, 33, 1--21.

\bibitem[\protect\citeauthoryear{Cai, Devanur, and Weinberg}{Cai, Devanur, and Weinberg}{2016}]{CDW16}
  \textsc{Cai, Y., Devanur, N. R., and Weinberg, S. M.} (2016): \enquote{A Duality Based Unified Approach to
Bayesian Mechanism Design,}
  \emph{Proceedings of the 48$^{th}$ Annual ACM Symposium on Theory of Computing}, 926–939.


  \bibitem[\protect\citeauthoryear{Carroll}{Carroll}{2017}]{Ca17}
  \textsc{Carroll, G.} (2017): \enquote{Robustness and Separation in Multidimensional Screening,}
  \emph{Econometrica}, 85, 453--488.


\bibitem[\protect\citeauthoryear{Chawla, Hartline, and Kleinberg}{2007}]{CHK07}
  \textsc{Chawla, S., J.~ Hartline, and R.~Kleinberg} (2007):
  \enquote{Algorithmic Pricing via Virtual Valuations,}
\emph{Proceedings of the 8th ACM conference on Electronic commerce,} 243--251.

\bibitem[\protect\citeauthoryear{Chen, He, Li, and Sun}{Chen
  et~al.}{2019}]{CHLS19}
\textsc{Chen, Y.-C., W.~He, J.~Li, and Y.~Sun} (2019): \enquote{Equivalence of
  Stochastic and Deterministic Mechanisms,} \emph{Econometrica}, 87,
  1367--1390.

\bibitem[\protect\citeauthoryear{Daskalakis, Deckelbaum, and Tzamos}{Daskalakis
  et~al.}{2017}]{DDT17}
\textsc{Daskalakis, C., A.~Deckelbaum, and C.~Tzamos} (2017): \enquote{Strong
  Duality for a Multiple-good Monopolist,} \emph{Econometrica}, 85, 735--767.

\bibitem[\protect\citeauthoryear{Devanur, Haghpanah, and Psomas}{Devanur
  et~al.}{2020}]{DHP20}
\textsc{Devanur, N.~R., N.~Haghpanah, and A.~Psomas} (2020): \enquote{Optimal
  Multi-unit Mechanisms with Private Demands,} \emph{Games and Economic
  Behavior}, 121, 482--505.

\bibitem[\protect\citeauthoryear{Dhangwatnotai, Roughgarden, and Yan}{2015}]{DRY15}
  \textsc{Dhangwatnotai, P., T.~Roughgarden, and Q.~Yan, } (2015):
\enquote{Revenue Maximization with a Single Sample},
   \emph{Games and Economic Behavior,} {91}, {318--333}.

\bibitem[\protect\citeauthoryear{Haghpanah and Hartline}{Haghpanah and
  Hartline}{2020}]{HH20}
\textsc{Haghpanah, N. and J.~D.~Hartline} (2020): \enquote{When is Pure Bundling
  Optimal?} \emph{forthcoming, Review of Economic Studies}.

  \bibitem[\protect\citeauthoryear{Hart and Nisan}{Hart and Nisan}{2017}]{HN17}
  \textsc{Hart, S. and N.~Nisan} (2017): \enquote{Approximate Revenue Maximization with Multiple Items,} \emph{Journal of
    Economic Theory}, 172, 313--347.

\bibitem[\protect\citeauthoryear{Hart and Nisan}{Hart and Nisan}{2019}]{HN19}
\textsc{Hart, S. and N.~Nisan} (2019): \enquote{Selling Multiple Correlated
  Goods: Revenue Maximization and Menu-size Complexity,} \emph{Journal of
  Economic Theory}, 183, 991--1029.

\bibitem[\protect\citeauthoryear{Hart and Reny}{Hart and Reny}{2015}]{HR15}
\textsc{Hart, S. and P.~J. Reny} (2015):
\enquote{Maximal Revenue with Multiple Goods: Nonmonotonicity and Other Observations,} \emph{Theoretical Economics}, 10, 893--922.

\bibitem[\protect\citeauthoryear{Hart and Reny}{Hart and Reny}{2019}]{HR19}
\textsc{Hart, S. and P.~J. Reny} (2019):
\enquote{The Better Half of Selling Separately,}
\emph{ACM Transactions on Economics and Computation (TEAC),} 7, 1--18.

\bibitem[\protect\citeauthoryear{Hartline  and Roughgarden}{2009}]{HR09}
\textsc{Hartline, J.~D.,  and T.Roughgarden} (2009):
 \enquote{Simple versus optimal mechanisms,}
\emph{Proceedings of the 10th ACM conference on Electronic Commerce,} {225--234}.

\bibitem[\protect\citeauthoryear{Kleiner and Manelli}{Kleiner and
  Manelli}{2019}]{KM19}
\textsc{Kleiner, A. and A.~Manelli} (2019): \enquote{Strong Duality in Monopoly
  Pricing,} \emph{Econometrica}, 87, 1391--1396.

 \bibitem[\protect\citeauthoryear{Laffont and Martimort}{Laffont and Martimort}{2002}]{LM02}
\textsc{Laffont, J.-J and D.~Martimort} (2002): \enquote{The Theory of Incentives,} \emph{Princeton University Press}.

\bibitem[\protect\citeauthoryear{Malakhov and Vohra}{Malakhov and
  Vohra}{2009}]{MV09}
\textsc{Malakhov, A. and R.~V. Vohra} (2009): \enquote{An Optimal Auction for
  Capacity Constrained Bidders: A Network Perspective,} \emph{Economic Theory},
  39, 113--128.

\bibitem[\protect\citeauthoryear{Manelli and Vincent}{Manelli and
  Vincent}{2006}]{MV06}
\textsc{Manelli, A.~M. and D.~R. Vincent} (2006): \enquote{Bundling as an
  Optimal Selling Mechanism for a Multiple-good Monopolist,} \emph{Journal of
  Economic Theory}, 127, 1--35.

\bibitem[\protect\citeauthoryear{Manelli and Vincent}{Manelli and
  Vincent}{2007}]{MV07}
---\hspace{-.1pt}---\hspace{-.1pt}--- (2007): \enquote{Multidimensional
  Mechanism Design: Revenue Maximization and the Multiple-good Monopoly,}
  \emph{Journal of Economic Theory}, 137, 153--185.

\bibitem[\protect\citeauthoryear{McAfee and McMillan}{McAfee and
  McMillan}{1988}]{MM88}
\textsc{McAfee, R.~P. and J.~McMillan} (1988): \enquote{Multidimensional
  Incentive Compatibility and Mechanism Design,} \emph{Journal of Economic
  Theory}, 46, 335--354.


  \bibitem[\protect\citeauthoryear{Menicucci, Hurkens, and Jeon}{2015}]{MHJ15}
  \textsc{Menicucci, D., S. Hurkens, and D. S., Jeon} (2015):
\enquote{On the Optimality of Pure Bundling for a Monopolist},
   \emph{Journal of Mathematical Economics} {60}, {33--42}.


\bibitem[\protect\citeauthoryear{Myerson}{Myerson}{1981}]{My81}
\textsc{Myerson, R.~B.} (1981): \enquote{Optimal Auction Design,}
  \emph{Mathematics of Operations Research}, 6, 58--73.

 \bibitem[\protect\citeauthoryear{Pavlov}{Pavlov}{2011a}]{Pa11a}
\textsc{Pavlov, G.} (2011a): \enquote{Optimal Mechanisms for Selling Two Goods,} \emph{The BE Journal of Theoretical Economics}, 11.

\bibitem[\protect\citeauthoryear{Pavlov}{Pavlov}{2011b}]{Pa11b}
\textsc{Pavlov, G.} (2011b): \enquote{A Property of Solutions to Linear Monopoly
  Problems,} \emph{The BE Journal of Theoretical Economics}, 11.

\bibitem[\protect\citeauthoryear{Pavlov}{Pavlov}{2020}]{Pa20}
---\hspace{-.1pt}---\hspace{-.1pt}--- (2020): \enquote{Selling Two Units of a
  Customizable Good,} Working paper, University of Western Ontario.

\bibitem[\protect\citeauthoryear{Pycia}{Pycia}{2006}]{Py06}
\textsc{Pycia, M.} (2006): \enquote{Stochastic vs Deterministic Mechanisms in
  Multidimensional Screening,} MPhil. Thesis, Mimeo available at
  http://pycia.bol.ucla.edu/pycia-multidimensional-screening.pdf.

\bibitem[\protect\citeauthoryear{Riley and Zeckhauser}{Riley and
  Zeckhauser}{1983}]{RZ83}
\textsc{Riley, J. and R.~Zeckhauser} (1983): \enquote{Optimal Selling
  Strategies: When to Haggle, When to Hold Firm,} \emph{The Quarterly Journal
  of Economics}, 98, 267--289.

\bibitem[\protect\citeauthoryear{Rochet}{Rochet}{1987}]{Ro87}
\textsc{Rochet, J.-C.} (1987): \enquote{A Necessary and Sufficient Condition
  for Rationalizability in a Quasi-linear Context,} \emph{Journal of
  Mathematical Economics}, 16, 191--200.

\bibitem[\protect\citeauthoryear{Rochet and Chon{\'e}}{Rochet and
  Chon{\'e}}{1998}]{RC98}
\textsc{Rochet, J.-C. and P.~Chon{\'e}} (1998): \enquote{Ironing, Sweeping, and
  Multidimensional Screening,} \emph{Econometrica}, 66, 783--826.



\bibitem[\protect\citeauthoryear{Shaked and Shanthikumar}{Shaked and
  Shanthikumar}{2007}]{SS07}
\textsc{Shaked, M. and J.~G. Shanthikumar} (2007): \emph{Stochastic Orders},
  Springer.

  \bibitem[\protect\citeauthoryear{Tang and Wang}{Tang and Wang}{2017}]{TW17}
  \textsc{Tang, P. and Z. Wang} (2017): \enquote{Optimal Mechanisms with Simple Menus,} \emph{Journal of Mathematical Economics}, 69, 54--70.



\bibitem[\protect\citeauthoryear{Thanassoulis}{Thanassoulis}{2004}]{Th04}
\textsc{Thanassoulis, J.} (2004): \enquote{Haggling over Substitutes,}
  \emph{Journal of Economic Theory}, 117, 217--245.

\bibitem[\protect\citeauthoryear{Wilson}{Wilson}{1993}]{Wi93}
\textsc{Wilson, R.~B.} (1993): \emph{Nonlinear Pricing}, Oxford University
  Press.

\end{thebibliography}


\end{document}